\documentclass[namedreferences]{solarphysics}
\usepackage[hyperref,optionalrh,solaromanenum]{spr-sola-addons} 
\usepackage{graphicx}                    
\usepackage{amssymb}                    
\usepackage{color}                       
\usepackage{breakurl}                         

\newcommand{\Bperp}{{\rm B}_\perp}
\newcommand{\Bpar}{{\rm B}_\parallel}
\newcommand{\Blos}{{\rm B}_{\rm los}}

\newcommand{\BB}{\mbox{\boldmath$B$}}
\newcommand{\Bxh}{B_x^h}
\newcommand{\Byh}{B_y^h}
\newcommand{\Bzh}{B_z^h}

\newcommand{\Bzi}{B_z^i}
\newcommand{\Bxi}{B_x^i}
\newcommand{\Byi}{B_y^i}

\newcommand{\Br}{B_r}
\newcommand{\Bth}{B_\theta}
\newcommand{\Bph}{B_\phi}
\newcommand{\synth}{{\tt SyntHIA}}
\newcommand{\vfisv}{{\tt VFISV}}
\newcommand{\vfisva}{{\tt VFISV\_ABGM}}
\newcommand{\unno}{{\tt UNNOFit}}
\newcommand{\merlin}{{\tt MERLIN}}
\newcommand{\asp}{{\tt ASP-ME}}

\begin{document}

\begin{article}

\begin{opening}

\title{On Identifying and Mitigating Bias in Inferred Measurements for 
Solar Vector Magnetic Field Data.}
%
\author[addressref={aff1,aff2},corref,email={leka@nwra.com}]{\inits{K.D.}\fnm{K.D.}~\lnm{Leka}\orcid{0000-0003-0026-931X}}
\author[addressref={aff1},email={wagneric@nwra.com}]{\inits{E.L.}\fnm{Eric L.}~\lnm{Wagner}\orcid{0000-0002-7709-723X}}
\author[addressref={aff3,aff4,aff5},email={a.b.g.marin@astro.uio.no}]{\inits{A. B.}\fnm{Ana Bel\'{e}n}~\lnm{Gri{\~n}{\'o}n-Mar{\'\i}n}\orcid{0000-0003-0026-931X}}
\author[addressref={aff6},email={V.Bommier@obspm.fr}]{\inits{V.}\fnm{V\'{e}ronique}~\lnm{Bommier}\orcid{0000-0002-6253-9170}}
\author[addressref={aff7},email={relh@umich.edu}]{\inits{R.}\fnm{Richard}~\lnm{Higgins}\orcid{0000-0002-6227-0773}}

%
\runningauthor{Leka et al.}
\runningtitle{Quantifying and Mitigating $\Bperp$ Bias}
\address[id=aff1]{NorthWest Research Associates, Boulder, Colorado, USA}
\address[id=aff2]{Institute for Space-Earth Environmental Research, Nagoya University, Nagoya, Aichi, Japan}
\address[id=aff3]{W. W. Hansen Experimental Physics Laboratory, Stanford University, Stanford, California, USA}
\address[id=aff4]{Institute of Theoretical Astrophysics, University of Oslo, Blindern, Oslo, Norway}
\address[id=aff5]{Rosseland Centre for Solar Physics, University of Oslo, Blindern, Oslo, Norway}
\address[id=aff6]{LESIA, Observatoire de Paris, Universit\'{e} PSL, Sorbonne Universit\'{e}, 
Universit\'{e} Paris Cit\'{e}, CNRS, Meudon, France}
\address[id=aff7]{University of Michigan, Ann Arbor, Michigan, USA}
\begin{abstract}
The problem of bias, meaning over- or underestimation, 
of the component perpendicular
to the line-of-sight (`$\Bperp$') in vector magnetic field maps is discussed.
Previous works on this topic have illustrated that the problem
exists; here we perform novel investigations to quantify the bias,
fully understand its source(s), and provide mitigation strategies.
First, we develop quantitative metrics to measure the $\Bperp$ bias
and quantify the effect in both local (physical) and native image-plane
components.  Second we test and evaluate different inversion options
and data sources, to systematically characterize the impacts of choices,
including explicitly accounting for the magnetic fill fraction $f\!f$.
Third we deploy a simple model to test how noise and different models
of the bias may manifest.  From these three investigations we find
that while the bias is dominantly present in under-resolved structures,
it is also present in strong-field pixel-filling structures.  Noise
in the magnetograms can exacerbate the problem, but it is not the primary cause.
We show
that fitting $f\!f$ explicitly provides significant mitigation, but that
other considerations such as choice of $\chi^2$ weights and optimization
algorithms can impact the results as well.  Finally,
we demonstrate a straightforward ``quick fix'' that can be applied {\it
post-facto} but prior to solving the $180^\circ$ ambiguity in $\Bperp$,
and which may be useful when global-scale structures are, {\it e.g.},
used for model boundary input.  The conclusions of this work support
the deployment of inversion codes that explicitly fit $f\!f$ or, as with
the new \synth\ neural-net, that are trained on data that did so.

\end{abstract}
\keywords{Instrumental Effects; Magnetic fields, Photosphere; Polarization, Optical}

\end{opening}
\section{Introduction}
\label{sec:intro} 

It is a challenging problem to infer the magnetic field strength and
direction in the solar photosphere as it threads a dynamic plasma
\citep[see][and references therein]{deltoroiniesta_ruizcobo_2016}.
The assumptions of a Milne-Eddington (ME)
atmosphere provide a good estimate of the average strength and direction
across the layers where magnetically-sensitive photospheric spectral
lines are formed \citep{tomography1}, especially in structures where a
single pixel-filling magnetic component is present or at least dominates.
Under-resolved multiple atmospheres (whether all magnetized or not)
contributing light to the resolution element result in polarimetric
signals that are intensity-weighted averages of the contributing
atmospheres \citep{mismas,magres} which rarely resemble expected ME
Stokes spectra (even without complications from unresolved velocity
components or gradients within the line-forming region).

The treatment of instrumental scattered light, and the approach used to estimate
relative contributions of magnetized {\it vs.} unmagnetized incoming
light (the magnetic ``fill fraction'', $f\!f$, or percentage of a
pixel filled by magnetic field) will influence the inferred
nature of magnetic structures, especially (but not solely) unresolved structures 
\citep{hifs,HSN_2004_multiline,ivm3,OrozcoSuarezKatsukawa2012,delToroIniesta_etal_2010,magres,Bommier2016,SainzDalda2017}.
The particulars of how inversion techniques are invoked by which
to infer the magnetic field from the spectropolarimetric data, even
``standard'' Milne-Eddington inversions, will influence the results.
The ``particulars'' can include the number and which spectral lines are
used, plus mundane-seeming choices of optimization algorithms, stopping
conditions, and any optimization weighting applied to $\chi^2$ calculation
\citep[see, {\it e.g.} ][for discussion]{hmi_invert,SainzDalda2017}.

As has been introduced at length in \citet{btrans_bias_1,Liu_etal_2022},
with vector polarimetry and inversions covering the full visible solar
disk from, {\it e.g.}, the Vector SpectroMagnetograph \citep[VSM][]{solis,VSM}
of the Synoptic Optical Long-term Investigations of the Sun 
facility ({\url{https://nso.edu/telescopes/nisp/solis/}}) and now the Helioseismic
and Magnetic Imager \citep[HMI][]{hmi,hmi_cal,hmi_invert,hmi_pipe} on the Solar Dynamics Observatory
\citep[{\it SDO};][]{sdo}, there is the capability of estimating the vector-components
of the photospheric magnetic field over large areas of the Sun.
Unfortunately, as demonstrated recently \citep{RudenkoDmitrienko2018,btrans_bias_1,Sun_etal_2021,Liu_etal_2022},
large-scale magnetic structures can present behavior that is physically
unexpected.  Specifically, a preferred direction in 
polar fields or a ``flip'' in the direction of the zonal-directed
(East/West) horizontal component of magnetic structures upon crossing the
central meridian.  Given that the
opposite signs of this bias are present in data from different facilities
\citep{btrans_bias_1,Sun_etal_2021}, the problem is solidified as originating
from instrumentation, data preparation, and/or inversion, rather than being
solar in origin.  We explore these options further, below.

What has not been emphasized yet is the impact on physical interpretation.
The over-estimation of the transverse component $\Bperp$ contributes to the 
inferred physical components $[\Bxh, \Byh, \Bzh]$, or heliographic / local
components of the field vector, in a 
non-linear way according to the inherent underlying structure
as well the viewing angle $\theta$ ({\it i.e.} the structure's location on the
solar disk). 

It has been proposed that the most susceptible features are those which are unresolved.  
As hypothesized in \citet{btrans_bias_1} and demonstrated in
\citet{Liu_etal_2022}, including the magnetic fill fraction $f\!f$
explicitly as part of a Milne-Eddington solution changes the 
inferred field vector, such that the bias may be mitigated
substantially.  This is not a new point
\citep{rmo87,ls90,ivm3,unnofit,delToroIniesta_etal_2010,SainzDalda2017}. 
Forcing the magnetic fill fraction $f\!f = 1.0$ for
unresolved structures is likely the dominant source of this bias; as shown
in \citet{Liu_etal_2022}, correcting this assumption by invoking a new
version of the Very Fast Inversion of Stokes Vector Milne-Eddington
approach \citep{vfisv,hmi_invert,abgm_etal_2021} can mitigate the
magnitude of the bias.  However, as also shown in \citep{btrans_bias_1},
the magnitude of the photon noise can contribute.  Spatial integrity,
field of view, and cadence are all drivers of instrument design, the
outcomes of which are as different as the questions they are optimized
to answer {\it c.f.} the {\it SDO}/HMI {\it vs.} the {\it Hinode}
SpectroPolarimeter \citep{hinode_sp_2013}, and we use both {\it SDO}/HMI and
{\it Hinode}/SP to investigate not just fill-fraction fitting, but noise,
instrument, data specifics, and inversion implementation.

Most importantly we present quantitative metrics by which to evaluate
the extent and direction of the bias\footnote{The authors became aware
of \citet{RudenkoDmitrienko2018} effectively after this paper was accepted.
We want to acknowledge some similarities of our approach to theirs,
state that the approaches were developed independently,
and direct interested readers to that paper for salient points.}
 and the contributing factors, and
finally demonstrate one ``quick and dirty'' correction which is shown
to mitigate (although not necessarily completely correct) the bias in
any data for which certain structures are observed and good coverage in 
observing angle is available.

We begin the methodology section (\S\,\ref{sec:methods}) with a
description of the data, target acquisition, and analysis methods
and evaluation metrics, plus a description of a simple model used
in the investigation.  This is followed by a summary of results
(\S\,\ref{sec:results}) and finally a demonstration of a rudimentary
correction approach (\S\,\ref{sec:qnd}) for when the option of
re-inverting the data is not available.

\section{Methodology}
\label{sec:methods}

In this section we present the algorithms for selecting features, 
the observational targets and analysis methodology.

The overall issue for {\it SDO}/HMI pipeline output is that the inferred component in
plane-of-sky is stronger than it ``should be'' compared to the line-of-sight
component, most apparently those with unresolved magnetic structure,
{\it i.e.} $f\!f < 1.0$.  For data from the Synoptic Optical Long-Term
Investigations of the Sun (SOLIS) Vector-SpectroMagnetograph \citep[VSM;][]{solis}, 
according to \citet{btrans_bias_1},
it is weaker than it ``should be''. We refer to this as the bias in $\Bperp$
but also refer to the image-plane (as returned from the inversion)
inclination $\gamma^i\in[0^\circ,180^\circ]$ where $[90^\circ]$ is in
the plane-of-the-sky.  In some cases, the ``polarity'' of $\Bpar =
\Blos$ is inconsequential or adds confusion, in which case we limit
the inclination, referred to as $|\gamma^i| \in [0^\circ,90^\circ]$.
Throughout, while we may discuss $f\!f$ as distinct from the field
magnitude $\BB$, it is their product that is evaluated.

\subsection{Target Features}
\label{sec:targets}

Solar plage is generally agreed to be comprised of kilo-Gauss
concentrations of predominantly radially-directed field; generally not 
strong enough to form a continuum depression (but magnetic and thermal
impacts are sufficient to produce continuum-intensity enhancements
when viewed at an angle), the small size of these
concentrations means they are rarely resolved with today's instruments.
Plage areas are not inherently ``weak field'' ({\it c.f.} the ``magnetic knots''
in \citet{RudenkoDmitrienko2018}).  The inability to 
resolve the structures leads to a canonical area-averaged (pixel-averaged) flux density
estimate for plage of $1 - \mathit{few} \times 100$\,Mx\,cm$^{-2}$,
depending on instrument specifics.  Plage is a wide-spread
source for ``global-scale'' solar magnetic field structure, and contributes strongly
to synoptic (or synodic) full-sun magnetic maps.

The overall approach we take relies on the analysis of structures on the
sun where the magnetic inclination is known, at least on a statistical
basis: plage regions, and the darkest umbrae of very stable sunspots.
These two targets, when carefully selected, should be dominated by
radially-directed field (for {\it SDO}/HMI spatial resolution and spectral sampling
and sensitivity).  It is not required that the inclination be {\it only}
radial, just that it be {\it dominated} by radially-directed field.

In cases where full-disk data are not available, there is an additional
assumption invoked, that these structures would display minimal large-scale
variation over the course of days, again on a statistical basis (for
appropriately stable sunspots).

Different between these two structures are the magnetic fill fraction,
with the darkest umbrae of sunspots presenting field-filled pixels ($f\!f =
1.0$), whereas plage should present un-resolved bundles of field within
field-free (or field-less) plasma ($f\!f < 1.0$).  Both plage and sunspots
generally present polarization signals well above the noise.
The two types of features appropriate for this analysis are selected thus: 
\begin{list}{\roman{enumi}:}{\usecounter{enumi}\setcounter{enumi}{0}\topsep0cm\leftmargin0.5cm} 
\item The very central darkest portions of sunspot umbrae in spots that are large 
and not evolving noticeably.  To select these areas, 
\begin{list}{\roman{enumi}.~\arabic{enumii}) }{\usecounter{enumii}\setcounter{enumii}{0}\itemsep0cm\leftmargin1.0cm}
\item Determine that the size, complexity, morphology and evolution of the target sunspot is large, 
nonexistent, round and steady, respectively.
\item Create four masks:
\begin{list}{$\circ$}{\leftmargin0.0cm\topsep0cm}
\item Mask 1, Continuum intensity: select the darkest points within sunspot umbrae.  Specifically, 
for {\it Hinode}/SP data: an $I_c / \rm{median}(I_c)$ limited-FOV image is smoothed with a 4-pixel boxcar, and
those pixels $ <0.4$ are selected; for {\it SDO}/HMI: using $I^*$ normalized continuum intensity from 
the {\tt hmi.Ic\_nolimbdark\_720s} data series \citep[see][for a description of the different
data series available from the {\it SDO}/HMI pipeline products]{hmi_pipe}, choose pixels with $I^* < 0.2$.
The resulting mask is then ``dilated'' by a $2\times 2$ box.
\item Mask 2, Field Strength: $f\!f\* BB > 200$; tests showed that this level consistently retains
plage areas but does not include noise across orbital variations in {\it SDO}/HMI data. This mask is then 
``eroded'' by a $2\times 2$ box to remove single-pixel detections.
\item Mask 3, Inclination: An image of the ``local'' or heliographic (or ``physical'') 
$|\gamma^h|$ is 
created, smoothed using a 4-pixel boxcar, and pixels $ < 30^\circ$ are identified; this mask
is then subjected to first being eroded and then grown both using a $2\times 2$ box.
\item Mask 4, Data Quality: for {\it SDO}/HMI, mask to include only those pixels with {\tt conf = 0.0}
from the {\tt confid\_map.fits} segment of {\tt hmi.ME\_720s\_fd10} series, indicating
no failures in data quality or inversion (convergence, {\it etc.}).  For {\it Hinode}/SP: only those
pixels with total polarization $P > 0.4$\%; those below are used for computing the 
non-magnetic spectral profiles.  For all: within $\mu = \cos(\theta) > 0.35$ (within $\approx 70^\circ$
from disk center).
\end{list} 
\item The four masks are added together and only pixels where all conditions are satisfied are selected.
\end{list} 
\item Plage areas are found either in the vicinity of a sunspot, or as
active-region remnants that can cover many degrees of the disk.  Arguably,
any individual plage element evolves, but statistically speaking their
distribution is not expected to differ due the hemisphere in which
they are located.  To select these areas,
\begin{list}{\roman{enumi}.~\arabic{enumii}:}{\usecounter{enumii}\setcounter{enumii}{0}\itemsep0cm\leftmargin1.0cm}
\item Again, construct a series of independent masks:
\begin{list}{$\circ$}{\leftmargin0.0cm\topsep0cm}
\item Mask 1, Continuum Intensity: where $I_c / \rm{median}(I_c) > 0.9$ ({\it Hinode}/SP) or $I^* > 0.9$ ({\it SDO}/HMI).
\item Mask 2, Field Strength: as per Mask 2 above, but additionally
grown with a $2\times 2$ box to ensure coverage of plage concentrations.
\item Mask 3, Inclination: same ask Mask 3, above.
\item Mask 4,  Data Quality: same as Mask 4, above.
\item Mask 5, extended sunspot / super-penumbral exclusion: using continuum intensity, select all
sunspot points by $I_c / \rm{median}(I_c) < 0.9$ ({\it Hinode}/SP) or $I^* < 0.9$
({\it SDO}/HMI). Erode the result using a $2\times 2$ box to remove single-point
detections, then grow with a large $r=25$-pixel circular template to extend spot
areas (including detected pores) beyond the horizontal-field super-penumbra.  
\end{list} 
\item Masks 1-4 are added together and only pixels where all conditions are satisfied are selected.
\item Mask 5 provides a negative-Boolean requirement, to remove not-dark but strong fields that can
confuse the analysis.
\end{list} 
\end{list} 

\noindent
The final masks included an additional identification of the polarity of $\Bzh$
and location on the disk (latitude, longitude).  Creating the plage masks
without Mask 3, meaning without a filter on the local-component inclination
angle, results in $\lesssim5^\circ$ difference in, {\it e.g.} the mean and
medians of the inclination angle distributions, by including non-radial
points.  As such, the interpretation of the bias may be systematically 
incorrect by a small amount, but avoiding having to resolve the $180^\circ$
ambiguity may be advantageous in some situations.

\subsection{Observational Data}
\label{sec:obs_data}

\subsubsection{{\it Hinode}/SpectroPolarimeter}
\label{sec:hinodesp}

\begin{figure}
\centerline{
\includegraphics[width=0.50\textwidth,clip, trim = 15mm 5mm 5mm 5mm, angle=0]{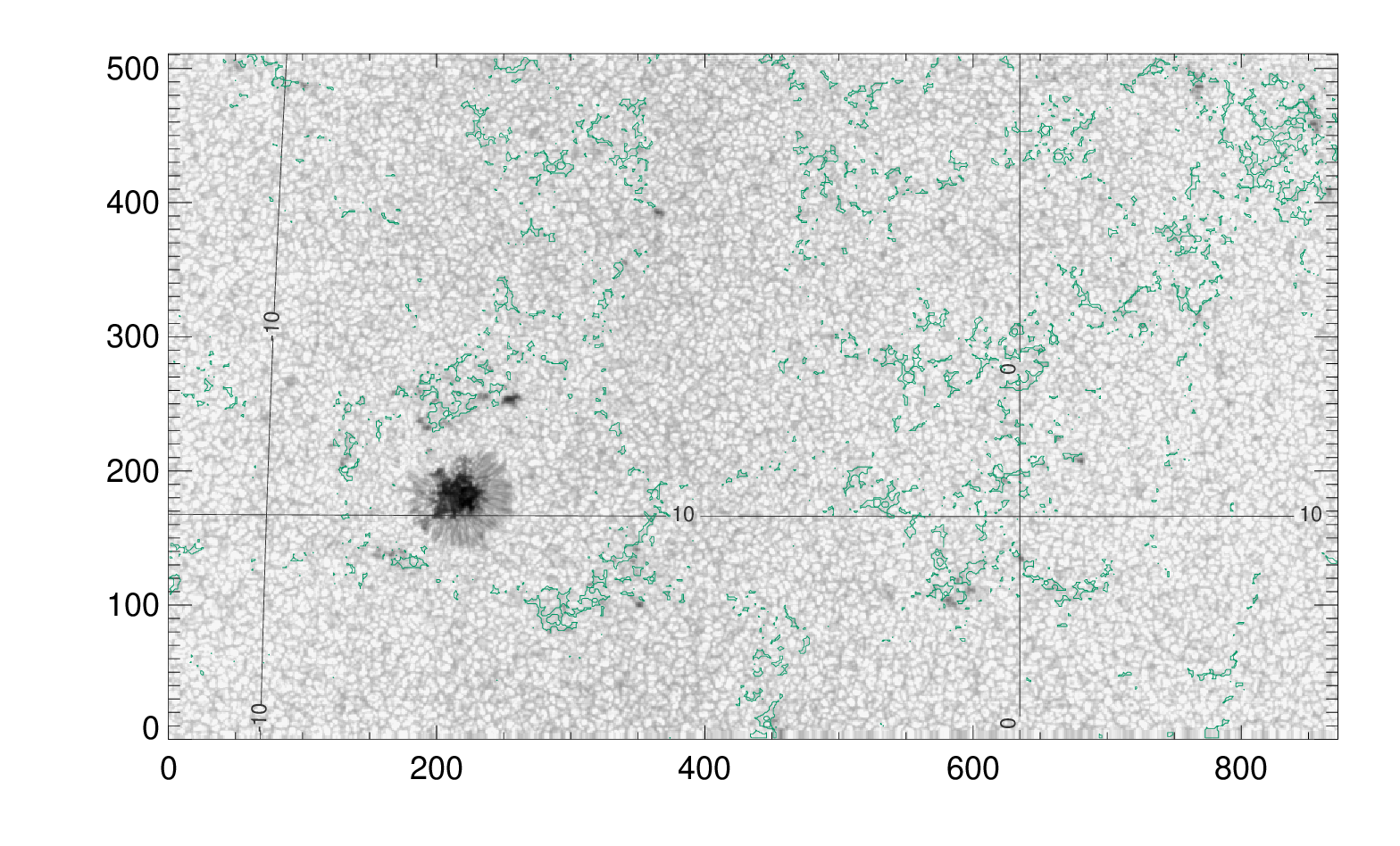}
\includegraphics[width=0.50\textwidth,clip, trim = 15mm 5mm 5mm 5mm, angle=0]{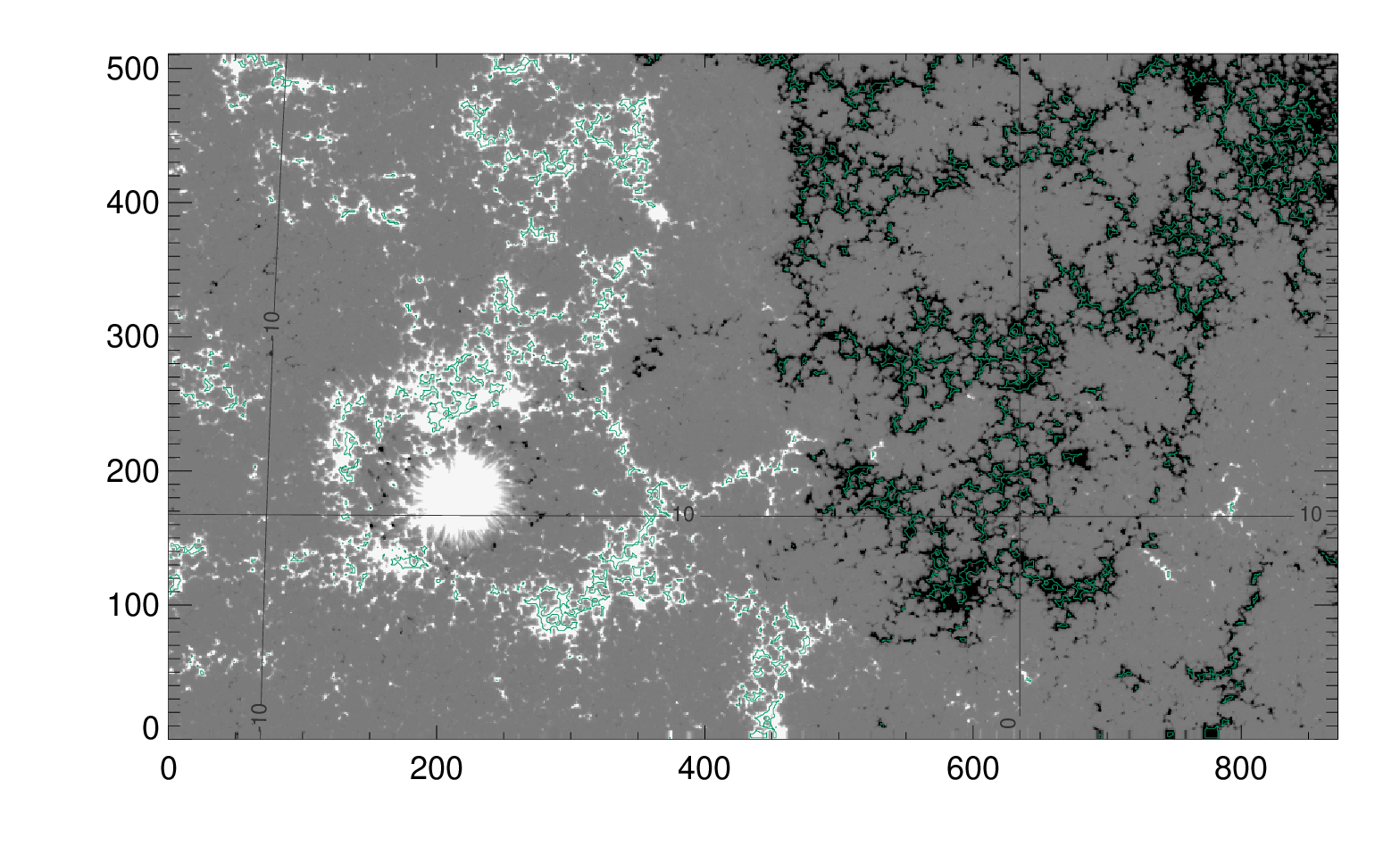}}
\caption{{\it Hinode}/SP ``normal'' scan of AR 12457, ID 20151124\_170005, showing (left) continuum (scaled to 1.1*median(Ic))
with latitude/longitude contours, and (right) $\Bzh$. Plage points are identified (green contours).
All target image axes are simply in pixel numbers.}
\label{fig:hinodefov}
\end{figure}

We chose {\it Hinode}/SpectroPolarimeter
\citep{hinode,hinode_sp,hinode_sot_cal,hinode_sp_2013} scans that follow
NOAA\,AR\,12457 (Table\,\ref{tbl:data}).  The target scans include a small
sunspot that does not fit our criteria for spot analysis, but does include
a well-developed and minimally-evolving plage area (Figure\,\ref{fig:hinodefov}).
Standard Level-1, the calibrated spectra, and Level-2, the \merlin Milne-Eddington 
inversion results, were retrieved; Level 2.1 data were not used
here, as we wanted to extend the disambiguation and ensure that it was consistent
across inversion experiments (see section~\ref{sec:inversions}). 

\subsubsection{{\it Solar Dynamics Observatory}/ Helioseismic and Magnetic Imager}

Data from the {\it Solar Dynamics Observatory}/ Helioseismic and Magnetic Imager 
\citep{sdo,hmi,hmi_invert,hmi_cal,hmi_pipe} provide the known problematic
data in this case, plus both time-series and full-disk data for testing inversion 
options and other mitigating procedures.  To match the 
{\it Hinode}/SP data of AR\,12457 we selected {\it SDO}/HMI segments close
to the mid-times of the {\it Hinode}/SP scans (see Table\,\ref{tbl:data})
and extracted coincident FOV patches from the full-disk data ({\it not} as defined by
the {\it SDO}/HMI Active Region Patch \#6124's bounding boxes).  The pipeline data included
series {\tt hmi.ME\_720s\_fd10} and {\tt hmi.B\_720s}, plus {\tt hmi.S\_720s}
for some tests of inversions, and {\tt hmi.Ic\_nolimbdark\_720s} for continuum-feature
identification.

\begin{figure}
\centerline{
\includegraphics[width=0.50\textwidth,clip, trim = 5mm 10mm 0mm 0mm, angle=0]{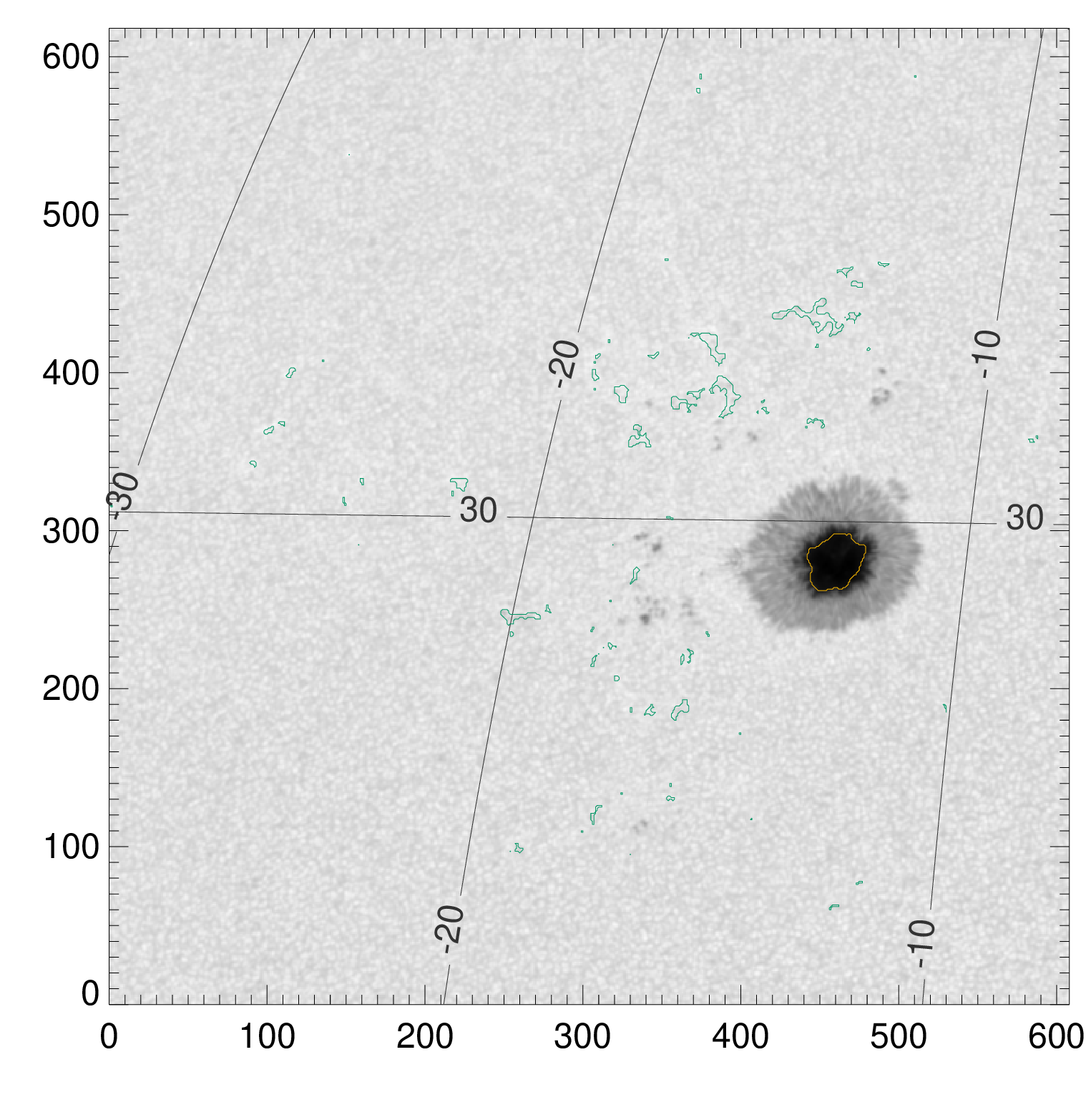}
\includegraphics[width=0.50\textwidth,clip, trim = 5mm 10mm 0mm 0mm, angle=0]{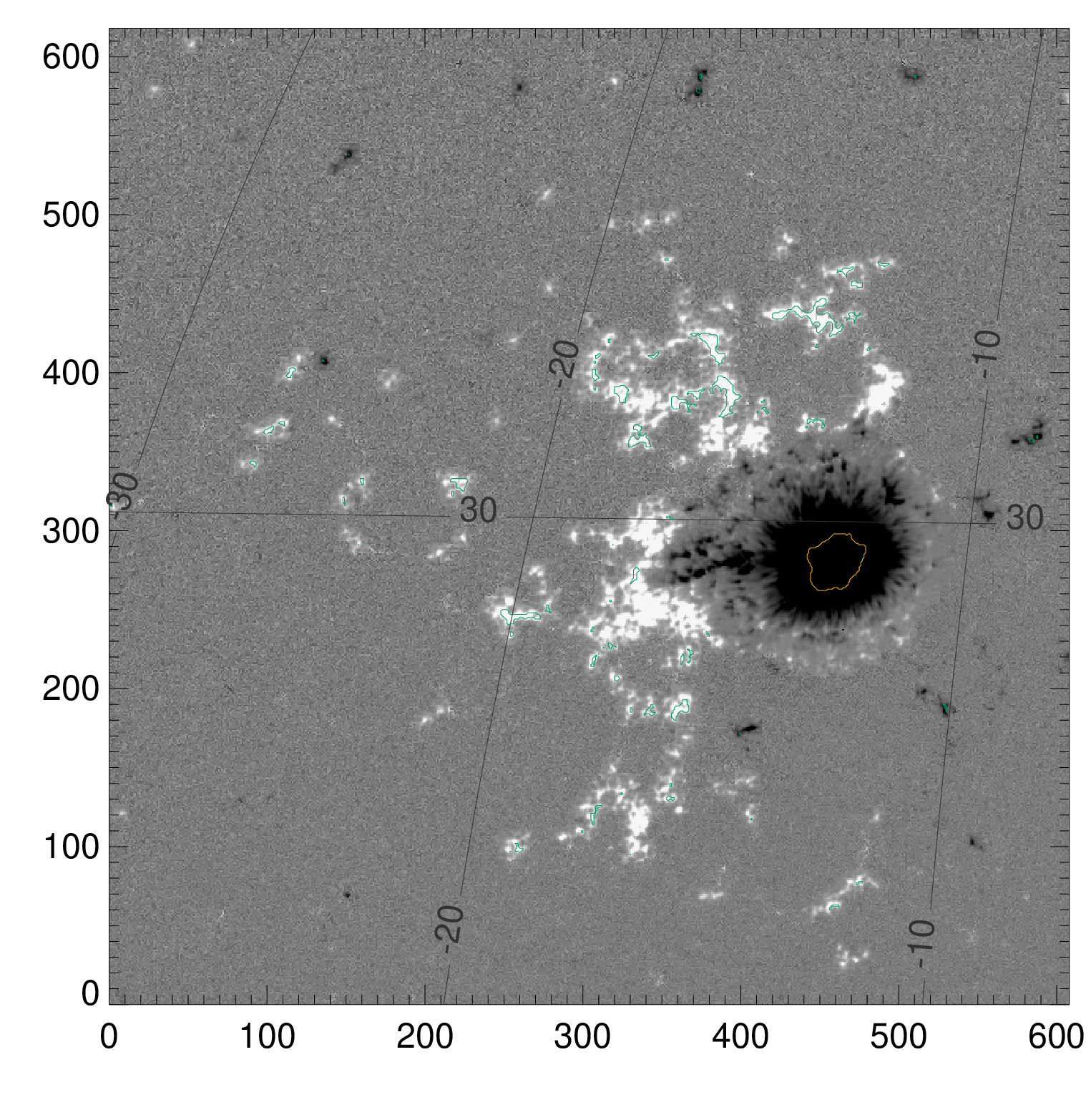}}
\caption{Target sunspot ``Spot 1'', NOAA\,AR\,11131 that transited the disk 2010.12.04 - 2010.12.12, at N30,
shown here at {\tt 2010.12.07\_07:24:00\_TAI}.  The surrounding plage is ``Plage 1''.
Left: Continuum, Stonyhurst grid, with axes indicating pixels.  
Right: $\Bzh$ from the Minimum-Energy disambiguation (see Section\,\ref{sec:ambig}),
same Stonyhurst grid, with contours of selected plage points (green) 
and center-spot radial-field area (yellow).}
\label{fig:hmi_spot1plage1}
\end{figure}

\begin{figure}
\centerline{
\includegraphics[width=0.50\textwidth,clip, trim = 5mm 10mm 5mm 0mm, angle=0]{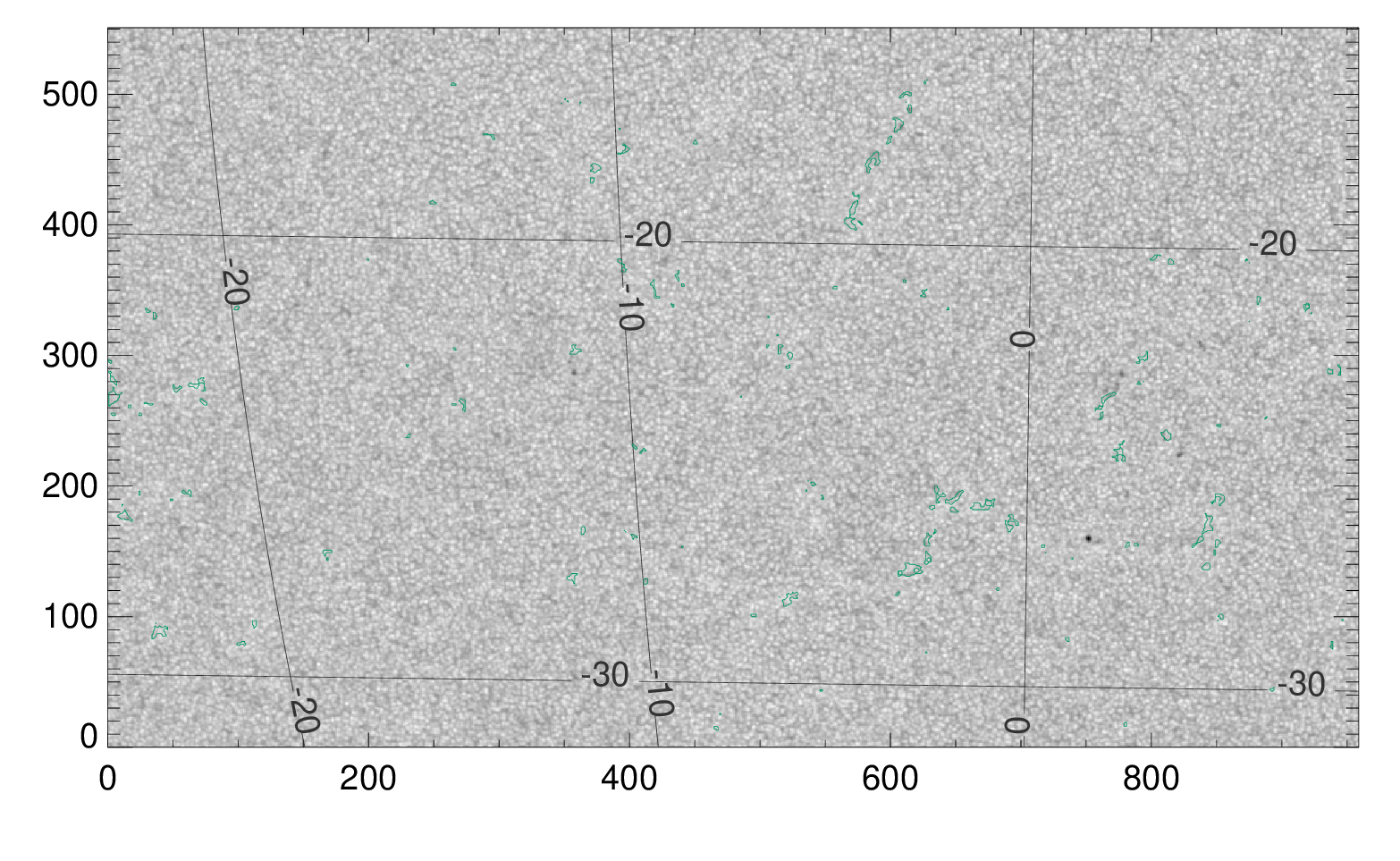}
\includegraphics[width=0.50\textwidth,clip, trim = 5mm 10mm 5mm 0mm, angle=0]{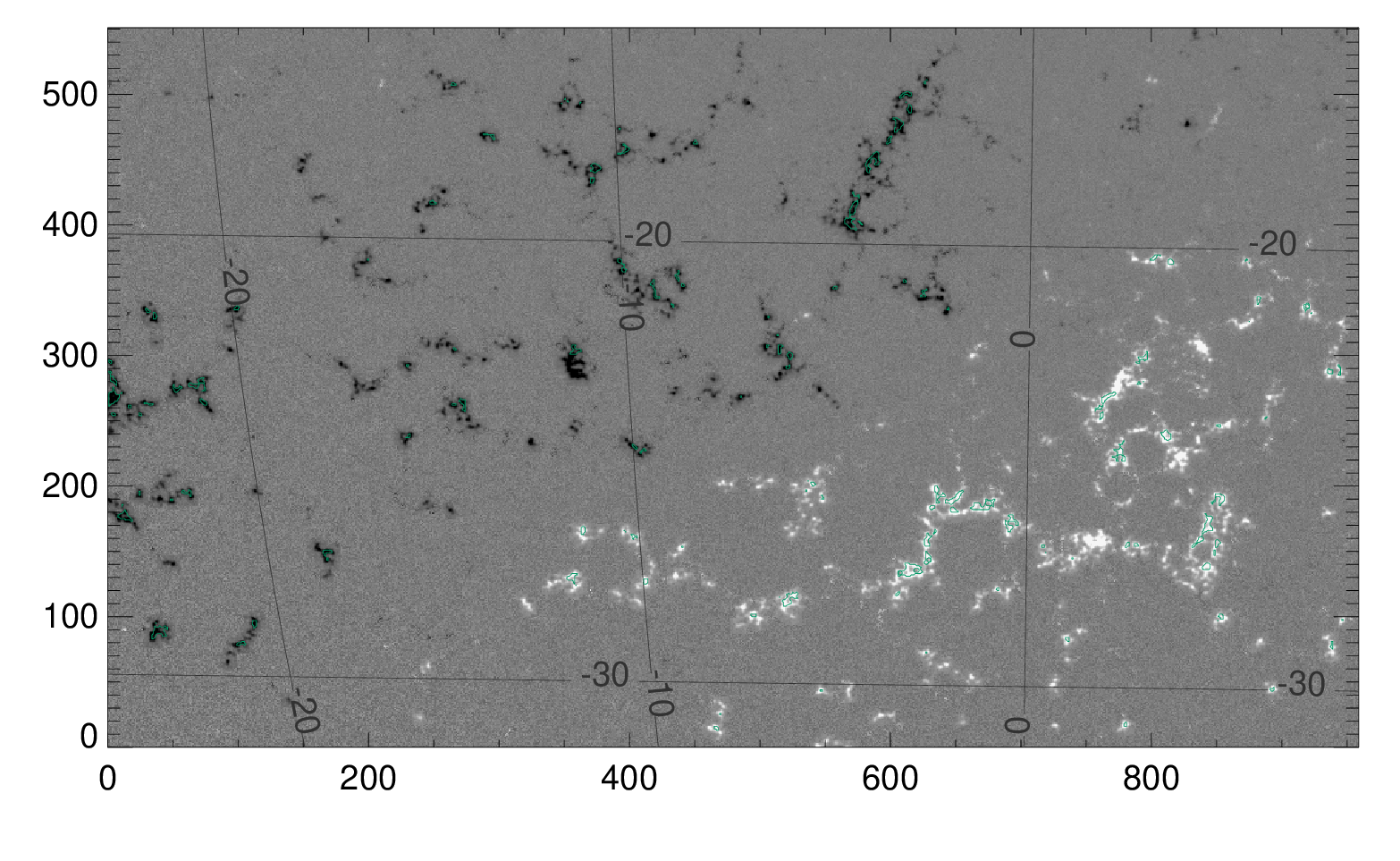}}
\caption{Same as Figure\,\ref{fig:hmi_spot1plage1} but for ``Plage 3'' that transited
2010.12.04 - 2010.12.12, almost directly south of NOAA\,AR\,11131 centered at S25, shown here at 
{\tt 2010.12.07\_20:12:00\_TAI}.}
\label{fig:hmi_plage3}
\end{figure}

\begin{figure}
\centerline{
\includegraphics[width=0.50\textwidth,clip, trim = 5mm 5mm 1mm 10mm, angle=0]{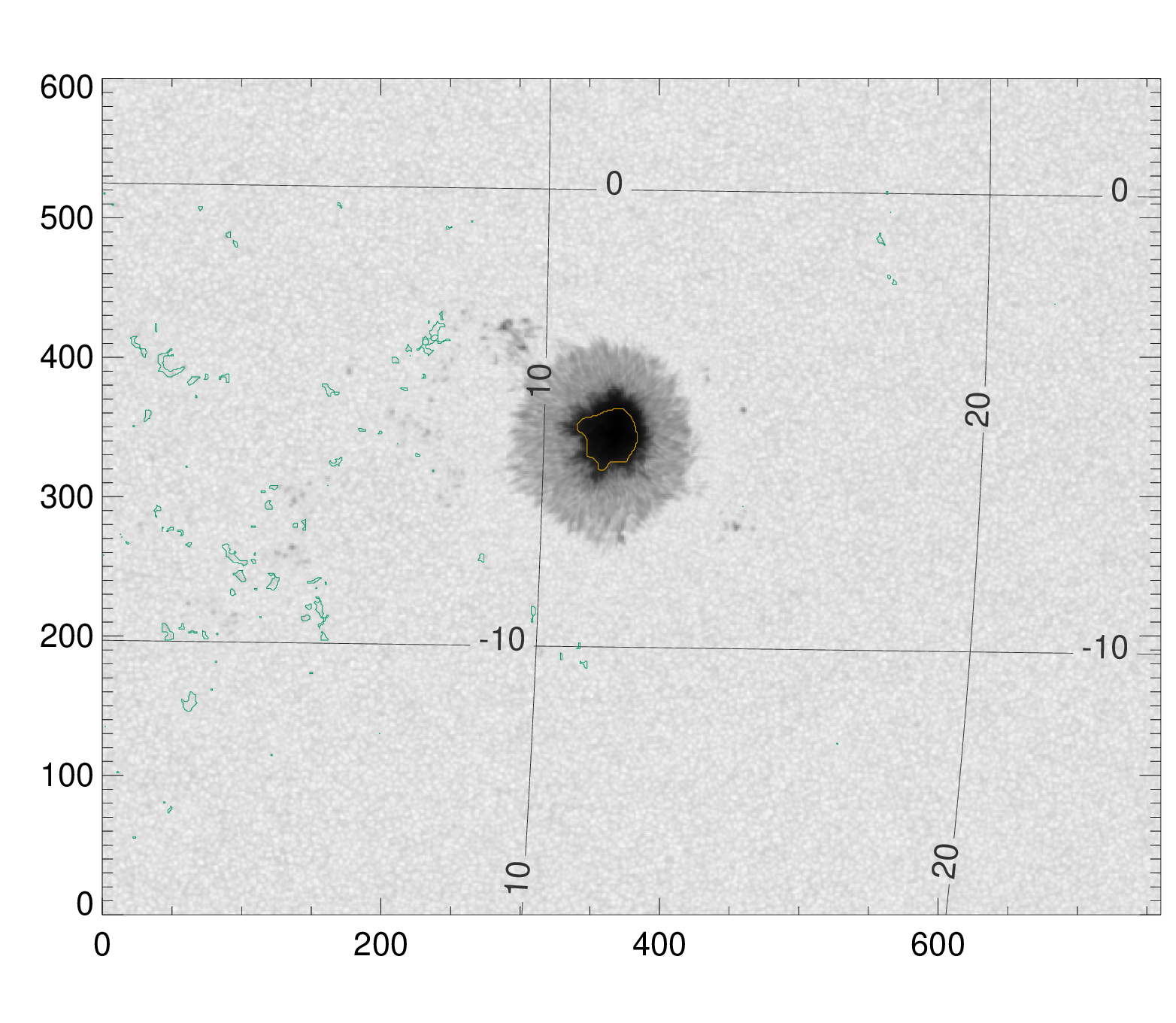}
\includegraphics[width=0.50\textwidth,clip, trim = 5mm 8mm 1mm 10mm, angle=0]{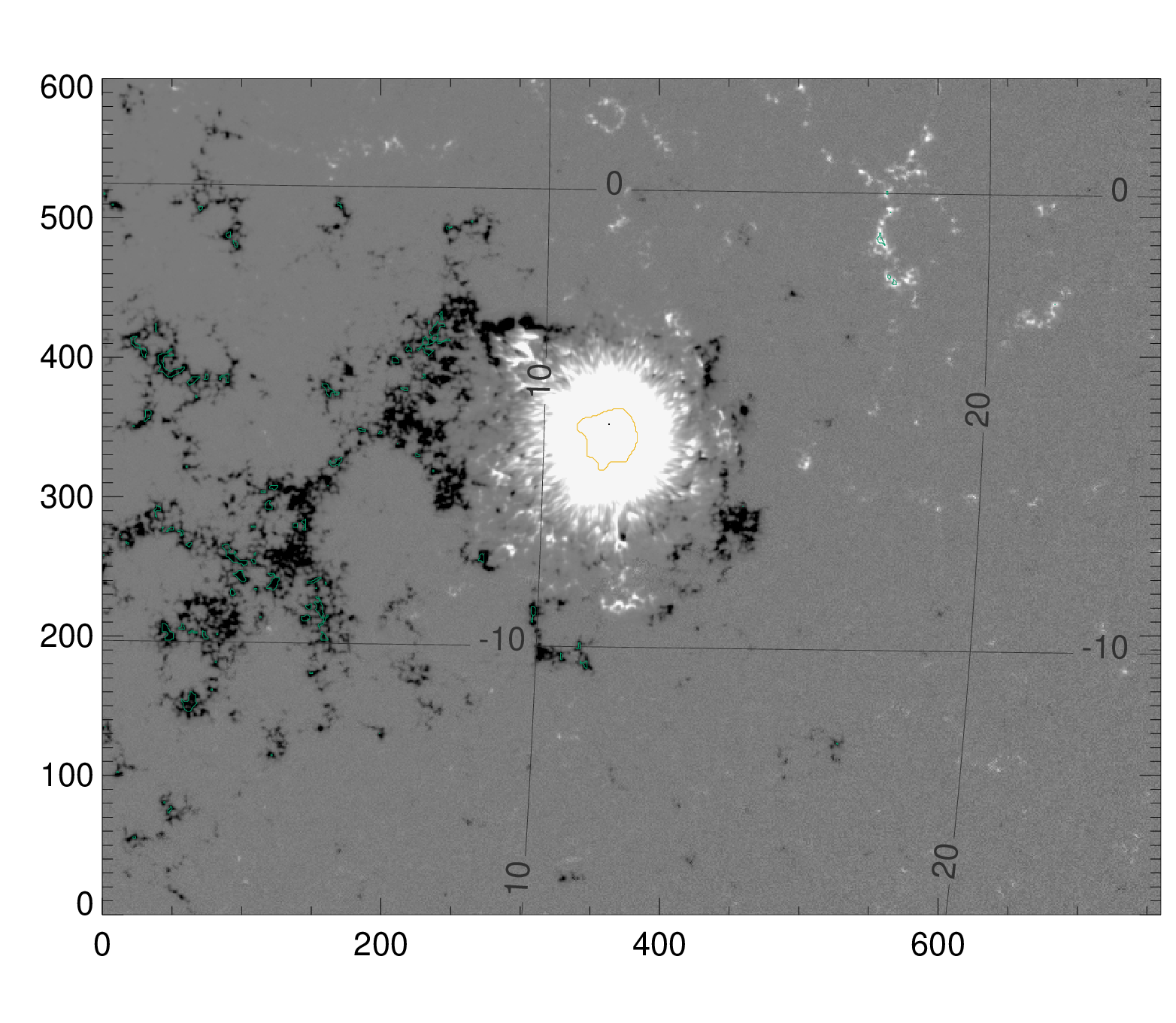}}
\caption{Same as Figure\,\ref{fig:hmi_spot1plage1} but for ``Spot 2'', NOAA\,AR\,12546,  and accompanying 
``Plage 2'' that transited the disk 2016.05.14 - 2016.05.25, shown here at {\tt 2016.05.21\_06:24:00\_TAI}.}
\label{fig:hmi_spot2plage2}
\end{figure}

\begin{figure}
\centerline{
\includegraphics[width=0.50\textwidth,clip, trim = 0mm 5mm 0mm 10mm, angle=0]{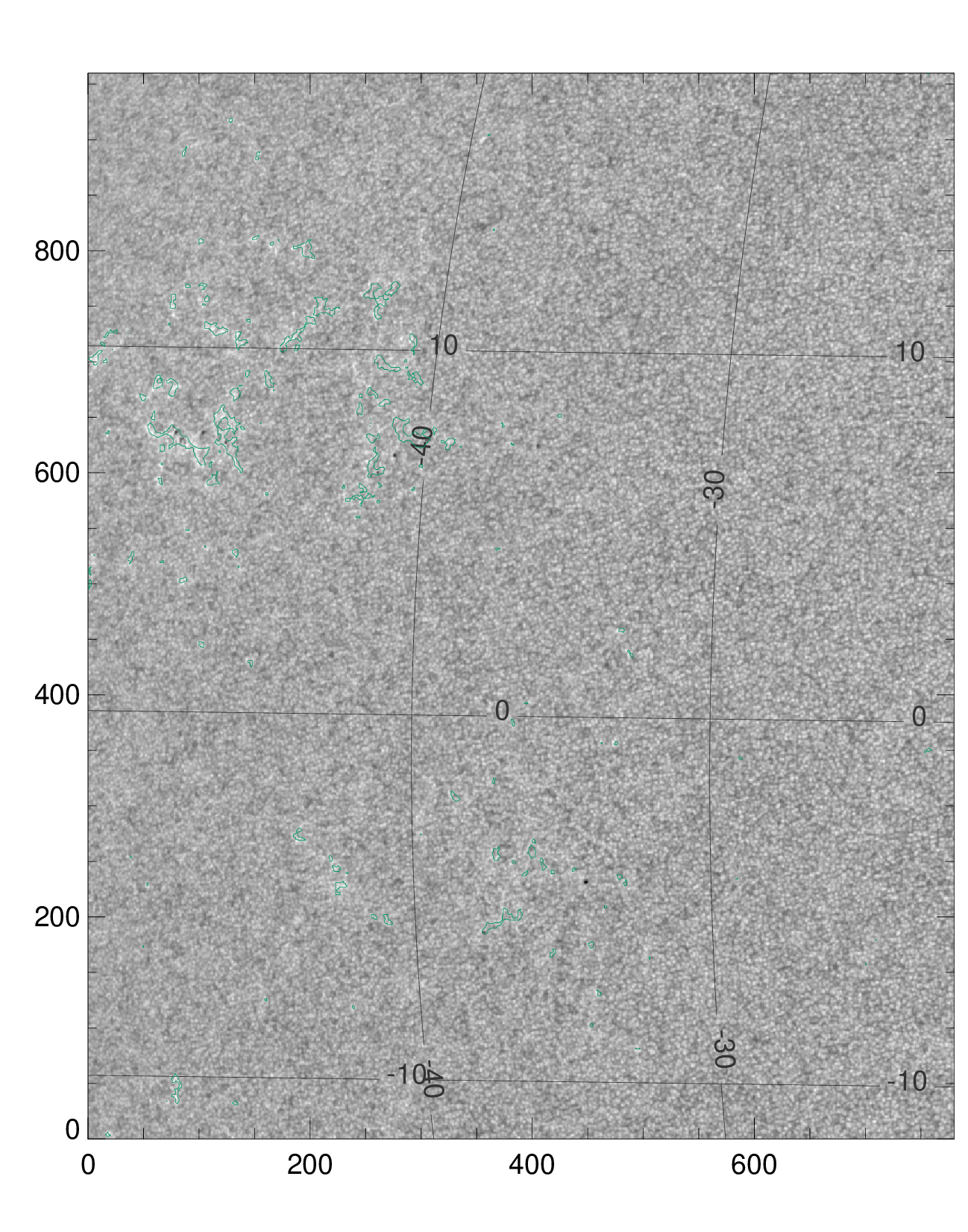}
\includegraphics[width=0.50\textwidth,clip, trim = 0mm 5mm 0mm 10mm, angle=0]{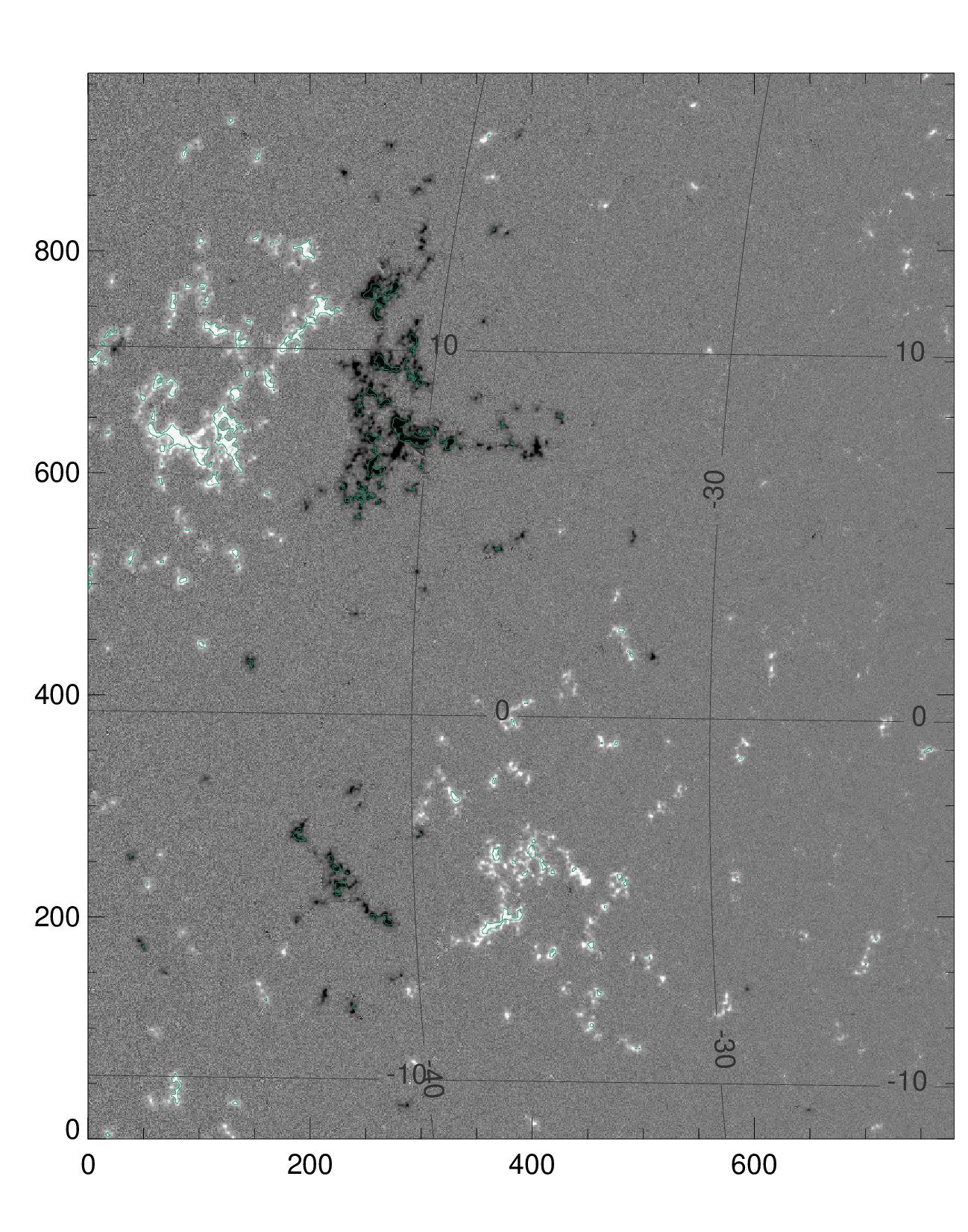}}
\caption{Same as Figure\,\ref{fig:hmi_spot1plage1} for target plage area ``Plage 4'' which transited
the disk 2016.05.13 - 2016.05.21, shown at {\tt 2016.05.14\_08:00:00\_TAI}.  Of note, this region was divided
into ``Plage 4 North'' and ``Plage 4 South'' for analysis.  Plage 4 South eventually 
produced sunspots and was labeled NOAA\,AR\,12547, but the sunspot points are excluded
for plage analysis.}
\label{fig:hmi_plage4}
\end{figure}

Two additional time-periods were identified for study, targeting
the presence of large, round, stable sunspots and well-distributed plage:
2010.12.04-2010.12.12 and 2016.05.13-2016.05.24 (Table\,\ref{tbl:data}).
Not all analysis methods were available for the earlier time-period, but
it was valuable to confirm behavior
with a second large sunspot and additional plage areas.
Custom-sized boxes were defined and extracted at 
the solar synodic rotation rate, to follow the targeted structures
across the solar disk.  In addition to the magnetic field-related 
segments, we utilized the confidence and the limb-darkened-corrected
continuum intensity (as mentioned above).
Context images for the targets selected from {\it SDO}/HMI time-series data are shown 
in Figures~\ref{fig:hmi_spot1plage1},~\ref{fig:hmi_plage3},~\ref{fig:hmi_spot2plage2},
~\ref{fig:hmi_plage4}.

\begin{table}
\caption{Data / Target Summary}
\begin{tabular}{lccl}
\hline
Target (Label) & Date/Time/ID & Location & Notes \\ \hline \hline
\multicolumn{4}{c}{Source: {\it Hinode}/SOT-SP \ \  Cadence: 1/day} \\ \hline
AR12467 & 20151122\_185049 & N12 E42 & HMI: 20151122\_190000 \\ 
$\prime\prime$ & 20151123\_181505 & N12 E29 & HMI: 20151123\_182400  \\ 
$\prime\prime$ & 20151124\_170005 & N12 E16 & HMI: 20151124\_171200 \\ 
$\prime\prime$ & 20151125\_154504 & N12 E03 & HMI: 20151125\_160000  \\
$\prime\prime$ & 20151126\_162004 & N12 W10 & HMI: 20151126\_163600 \\
$\prime\prime$ & 20151127\_151505 & N12 W23 & HMI: 20151127\_152400 \\ \hline
\multicolumn{4}{c}{Source: {\it SDO}/HMI \ \ Cadence: 96\,m} \\ \hline
Spot 1 (Sp1) & 2010.12.04 - 2010.12.12 & N32 & NOAA AR 11131 \\
Plage 1 (Pl1) & $\prime\prime$ & $\prime\prime$ & Surrounding Spot 1 \\
Plage 3 (Pl3) & $\prime\prime$ & S23 & broad in longitude extent \\ 
Spot 2 (Sp2) & 2016.05.17 -- 2016.05.25 & S07 & NOAA AR 12546 \\
Plage 2 (Pl2) & $\prime\prime$ & $\prime\prime$ & Surrounding Spot 2 \\
Plage 4 North (Pl4N) & 2016.05.13 -- 2016.05.21 & N07 &  \\
Plage 4 South (Pl4S) & $\prime\prime$ & S06  & includes AR 12547 \\ \hline
\end{tabular}
\label{tbl:data}
\end{table}

\subsubsection{Inversion Options}
\label{sec:inversions}

A number of different inversion options were evaluated, comprising two 
different groups of tests.  

The first group used the Milne-Eddington inversion code developed for
use with the NCAR/High Altitude Observatory Advanced Stokes Polarimeter
\citep[labeled \asp; ][]{sl87,ls90,bridges} on {\it Hinode}/SP
Level-1 data to systematically evaluate the impact of different
implementation options.  The options investigated include (see summary
in Table~\ref{tbl:Hinodegrid}):
\begin{list}{\arabic{enumi})}{\topsep0cm\leftmargin0.5cm\itemsep0cm\usecounter{enumi}
\setcounter{enumi}{0} }
\item optimization scheme: a genetic algorithm or least-squares minimization
to obtain the global minimum of $\chi^2$ fit between model atmosphere and observed spectra.
\item Fitting both $630.15, 630.25$\,nm lines or just the $630.25$\,nm.  Different $g_L$ 
factors imply different sensitivity to $\BB$ but not $f\!f$, so multiple lines provide
additional constraints.
\item Explicitly fitting for $f\!f$ {\it vs.} explicitly setting $f\!f = 1.0$.
\item Weights assigned to $[I,~Q,~U,~V]$: since the magnitude of the polarization
signals in $[V]$ and $[Q,~U]$ are of order $d {\rm I} / d\lambda$ and 
$d^2{\rm I} / d\lambda ^2$ respectively and photon noise scales accordingly, 
in order to insure that the $I$ does not dominate the $\chi^2$ of the 
evaluation functions, weights $[w_I, w_Q = w_U, w_V]$ are usually a parameter
supplied for the $\chi^2$ calculation.  We quote here $W_{\rm S}^{\rm eff} = W_s / \sigma_{\rm S}$
Care must be taken when comparing codes and their parameter settings, as some request
$w_X$ while some refer to $w_X^2$. 
\end{list}
\noindent
The second group compares different inversion codes that may differ in a variety
of ways.  The different codes are tested on one or both of {\it SDO}/HMI {\tt hmi.S\_720s} polarization
images or {\it Hinode}/SP Level-1 spectra, as listed in Table~\ref{tbl:HMIgrid}.
The different systems tested include:
\begin{list}{$\bullet$}{\topsep0cm\leftmargin0.5cm\itemsep0cm}
\item The pipeline {\it Hinode}/SP Level-2 output from the \merlin\ Milne-Eddington 
code (heredity from \asp). 
\item The pipeline output from the Very Fast Inversion of Stokes Vectors 
\citep[\vfisv][]{vfisv,hmi_invert} from the {\tt hmi.ME\_720s\_fd10} series.
\item The \unno\ code \citep{unnofit} was applied only to the
AR\,12457 targets, but with both {\it Hinode}/SP Level-1 data and {\it SDO}/HMI
{\tt hmi.S\_720s} data as input.  \unno\ introduced the magnetic
filling factor as a free parameter of the Milne-Eddington inversion. All
free parameters describing the non-magnetic part of the atmosphere are
set to equal those of the magnetic part, except for the magnetic field
vector itself. This procedure was recently implemented in a new \vfisv\
inversion (see next point), with a slight difference in the series
of the eight other parameters, where \unno\ also determines the
Voigt parameter $a$ and eliminates the source function at the photosphere
base by normalization.  \unno\ was the first code that explicitly
allowed a variable atmosphere for the non-magnetic component, although
it is set to be identical to the atmosphere of the magnetic component.
However, \citet{unnofit} also showed that nine parameters can exceed the
information content of the spectra as needed for a successful inversion,
and that only the pixel-averaged average magnetic field strength $f\!f*B$
is finally determined.
\item A new version of the Very Fast Inversion of Stokes Vector
\citep[\vfisva;][]{vfisv,abgm_etal_2021} was developed for implementation 
with {\it SDO}/HMI data to explicitly include a fit for the fill fraction.  Similar in the 
approach to \unno\ (see above), it should be noted that
the Voigt parameter is fixed at $a = 0.5$  and the source function at the
base of the photosphere is a fitted parameter.  The
results from this inversion have been shown to mitigate the bias \citet{abgm_etal_2021,Liu_etal_2022}.
One difference between the \vfisva\ data used here and that presented in
\citet{abgm_etal_2021,Liu_etal_2022} is that we used no polarization threshold for 
the inversion whereas a threshold of 0.25\% was used in those cited works.  We 
removed the polarization threshold in order to ensure no discontinuity near weak-signal plage areas.  
No additional scattered light correction was performed.
\item The \synth\ approach is not really an inversion {\it per se},
it is a Convolutional Neural Network that has been trained on the {\it Hinode}/SP
Level-2 output with {\it SDO}/HMI {\tt hmi.S\_720s} polarization images as input \citep{synode,synthia}.
The full-disk results provide field and fill-fraction separately, with
an overall fidelity closer to {\it Hinode}/SP pipeline output than
{\it SDO}/HMI pipeline output.
\end{list}

\noindent
We additionally test the question of polarization noise.  
As briefly demonstrated in \citep{btrans_bias_1}, noise in the linear polarization
$[Q,\,U]$ signals contribute to the bias.  We conduct tests comparing
inversions using input spectra from the {\it SDO}/HMI {\tt hmi.S\_720s} series to those using input spectra from 
the {\tt hmi.S\_5760s} series, which are 96\,m integrated Stokes spectra, and demonstrably lower
in random (photon) noise.

Not all possible permutations were executed, but a sufficient number with 
sufficient range of options so as to quantitatively judge the impact
of the different approaches to the bias.  The tests are summarized in 
Tables~\ref{tbl:Hinodegrid},~\ref{tbl:HMIgrid}; please note the method monikers
will be used in later discussion and figures.

\begin{table}
\caption{Tests Run on {\it Hinode}/SP target AR 12457}
\begin{tabular}{llllll}     
\hline
 Line(s) & $f\!f$ fit& Optimization & $W_{\rm S}^{\rm eff}$ & Moniker & Notes \\ 
 (nm)  &  explicitly? & method & [I,\,Q\&U,\,V] & &  \\ \hline
\multicolumn{6}{c}{Data Source: {\it Hinode}/SP; Inversion: \asp\footnotemark[1]} \\ \hline
630.15, .25 & $\surd$ & Genetic & [1,10,3.2] & A-DEF & ``Default'' \\
630.15, .25 & $\surd$ & Genetic & [1,31.2,3.2] & Aw1 & weights-test \#1  \\
630.15, .25 & $\surd$ & Least-Squares & [1,31.2,3.2] & ALSw1 &  \\
630.15, .25 & $\surd$ & Genetic & [1,10,2.2] & Aw2 & weights-test \#2  \\
630.15, .25 & $\surd$ & Least-Squares & [1,10,2.2] & ALSw2 &  \\
630.15, .25 & $\surd$ & Genetic & [1,1,1] & Aw3 & weights-test \#3 \\
630.15, .25 & $\surd$ & Genetic & [1,3.2,3.2] & Aw4 & weights-test \#4 \\
630.15, .25 & $\surd$ & Least-Squares & [1,10,3.2] & ALS & \\
630.25 & $\surd$ & Genetic & [1,10,3.2] & A1 & \\
630.25 & $\surd$ & Genetic & [1,31.2,3.2] & A1w1 & \\
630.25 & $\surd$ & Genetic & [1,10,2.2] & A1w2 & \\
630.25 & $\surd$ & Least-Squares & [1,10,3.2] & A1LS & \\
630.25 & X & Genetic & [1,10,2.2] & A1NFw2 & \\
630.25 & X & Genetic & [1,10,3.2] & A1NF & \\
630.25 & X & Least-Squares & [1,10,3.2] & A1NFLS & \\
630.15, .25 & X & Genetic &  [1,10,2.2] & ANFw2 & \\
630.15, .25 & X & Genetic & [1,10,3.2] & ANF & \\
630.15, .25 & X & Least-Squares  & [1,10,3.2] & ANFLS \\ \hline
\multicolumn{6}{c}{Data Source: {\it Hinode}/SP; Inversion: OTHER} \\ \hline
630.[15,.25] & $\surd$ & LM\footnotemark &  [0.01,1,0.1] & \merlin\ & L2 pipeline\footnotemark \\
630.15& $\surd$ & LM & [1,1,1] & UnnoFit\_6301 & \unno\footnotemark \\
630.25& $\surd$ & LM & [1,1,1] & UnnoFit\_6302 & \unno \\ \hline
\multicolumn{6}{c}{Data Source: {\it SDO}/HMI; Inversion: \vfisv} \\ \hline
617.3  & X & LM & [1,3,2] & PIPE & {\tt hmi.ME\_720s\_fd10}\footnotemark\\
617.3  & $\surd$ &  LM	& [1,3,2] & ABGM & {\tt su\_abgm}\footnotemark \\ \hline
\multicolumn{6}{c}{Data Source: {\it SDO}/HMI; Inversion: OTHER} \\ \hline
617.3 & $\surd$ & CNN & n/a & \synth\ & CNN\footnotemark\\
617.3 & $\surd$ & LM & [1,1,1] & UnnoFit\_HMI & \unno \\
\hline
\end{tabular}
1: \citet{sl87,bridges} \\
2:  Levenberg-Marquardt (LM) minimization algorithm \citep[see][]{NR} \\
3: {\url{https://sot.lmsal.com/data/sot/level2d/}} \\
4: \citet{unnofit} \\
5: \citet{hmi_invert} \\
6: \citet{abgm_etal_2021}\\
7: \citet{synode,synthia}\\
\label{tbl:Hinodegrid}
\end{table}

\begin{table}
\caption{Tests Run on {\it SDO}/HMI Multi-Day TimeSeries Targets}
\begin{tabular}{llll}     
\hline
  $f\!f$ Fit?  & Optimization & Moniker & Notes \\ \hline \hline
\multicolumn{4}{c}{2010.12.04\_05:48:00 -- 2010.12.12\_04:12:00: Spot \#1, Plage \#1, Plage \#3} \\ \hline
 X & LM &  PIPE & {\it SDO}/HMI pipeline \\ 
 $\surd$ & LM & ABGM & {\tt su\_abgm} \\ 
 $\surd$ & CNN &  \synth\ & \synth\ CNN\\ \hline
\multicolumn{4}{c}{2016.05.13\_20:48:00 -- 2016.05.24\_19:12:00: Spot \#2, Plage \#2, Plage \#4} \\ \hline
 X & LM & PIPE  & {\it SDO}/HMI pipeline \\
 X & LM & PIPE\_5760 & Input: {\tt hmi.S\_5760s} \\
 $\surd$ & LM & ABGM & {\tt su\_abgm} \\
 $\surd$ & LM & ABGM\_5760 & Input: {\tt hmi.S\_5760s}, {\tt su\_abgm} \\
 $\surd$ & CNN & \synth\ & \synth\ CNN\\ \hline
\end{tabular}
\label{tbl:HMIgrid}
\end{table}

\subsubsection{Disambiguation}
\label{sec:ambig}

The data from all inversions of both {\it Hinode}/SP and
{\it SDO}/HMI data are inherently $180^\circ$ ambiguous in
the plane-of-the-sky component.  All inverted data that were
compared using the local components were disambiguated
using the ``minimum-energy'' method \citep[ME0;][available at {\tt
www.nwra.com/AMBIG}]{metcalf94,hinode2ambig,hmi_pipe} with a cooling
schedule that generally matched both {\it SDO}/HMI and the {\it
Hinode}/SP~Level2.1 data products: {\tt tfactr}=0.98, {\tt neq}=100.
The primary difference in how {\tt ME0} was called relative to the {\it SDO}/HMI pipeline
concerns a lower
{\tt [athresh, bthres]}=[95,100] to capture more near-plage areas,
and deploying the spherical option for the larger-FOV {\it SDO}/HMI data.
The local components were computed then using planar or
spherical geometry, accordingly.

\subsection{Model Data}
\label{sec:model_data}

\begin{figure}
\centerline{
\includegraphics[width=0.6\textwidth,clip, trim = 5mm 0mm 8mm 2mm, angle=0]{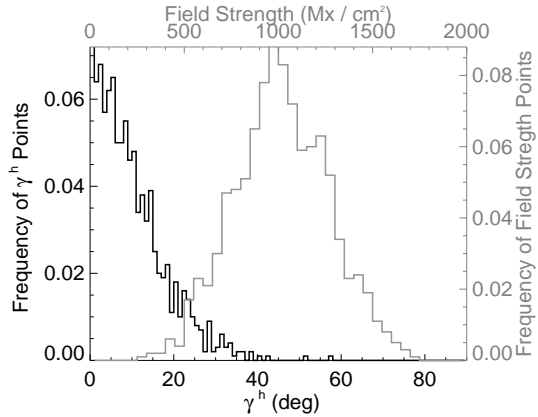}}
\caption{Distributions of heliographic or local inclination angle $\gamma^h$ (black) and
field strength (grey) for the ``plage-like'' distributions in the model.
$\gamma=0$ is radially-directed (opposite gravity).
These 1000-point samples were then placed across a variety of viewing angles.}
\label{fig:model_intro}
\end{figure}

To complement the analysis of observational data, model data were
constructed with which to test the hypotheses and the mitigation
strategies.  On a latitude / longitude grid ranging in both directions
$\pm 80^\circ$ in $2.5^\circ$ increments, 1,000-point samples were
generated to mimic expected distributions of plage: the radial component
is a normally-distributed random sample centered at 1,000G with a standard
deviation of 250G, the horizontal component was a normal distribution,
standard deviation 200G; the azimuthal angle was a uniform distribution
across $2\pi$ (Figure\,\ref{fig:model_intro}).  Only $\Bzh > 0$ 
positively-polarity ``plage'' was considered here.

From this grid, coordinate transforms re-create the image-plane components
$|\BB|, \gamma^i, \alpha^i$ or alternatively $\Bxi, \Byi, \Bzi$, the last
being $\Bpar$ and $\Bperp = \sqrt{{\Bxi}^2 + {\Byi}^2}$.
Noise was optionally added at `low', 'medium', and 'high' levels: 
a normal-distribution
random sample with $\sigma = [5, 10, 50]$G for $\Bpar$ was paired 
with a normal distribution with $\sigma = [100, 200, 300]$G for $\Bperp$, although
the absolute value was added (since the effective photon noise for
total linear polarization is positive-definite, see discussion in
\citet{btrans_bias_1}).  The noise in azimuthal angle was a normal
distribution with standard deviation approximately $[2, 4, 6]^\circ$.
In an attempt to mimic the noise maps of {\it SDO}/HMI \citep[see ][their Figure 7]{hmi_pipe},
we tested adding noise at the levels listed above but scaled as $1.0/\sqrt{\mu}$
such that the noise increases near the limb, 
but this made little difference in the final outcomes.
No attempt was made to model the impact of unresolved field or incorrect
fill factor \citep[although see ][]{magres}.

\subsection{Analysis and Metrics}

Simply detecting a ``sign flip'' in a re-binned image of 
a local heliographic field component as a plage area transits the central meridian
is a useful initial diagnostic \citep{btrans_bias_1,Liu_etal_2022}.  
However, it is difficult to quantitatively evaluate the 
bias in this manner ({\it c.f.} \citep{RudenkoDmitrienko2018}), or 
determine its full impact.

We develop quantitative metrics based on the believed physical characteristics
of solar structures in order to evaluate the bias.  
In some cases, temporal sampling is mapped to 
spatial sampling, again relying on the physical and statistical 
qualities of plage in that the distributions of the underlying
fields are not expected to evolve significantly over the course 
of a few days.

\subsubsection{Distributions of $\Bxh$, $\Byh$, $\Bzh$ with Central Meridian Distance and Observing Angle}
\label{sec:helio_components_distrib}

\begin{figure}[t]
\includegraphics[width=1.0\textwidth,clip, trim = 0mm 0mm 0mm 0mm, angle=0]{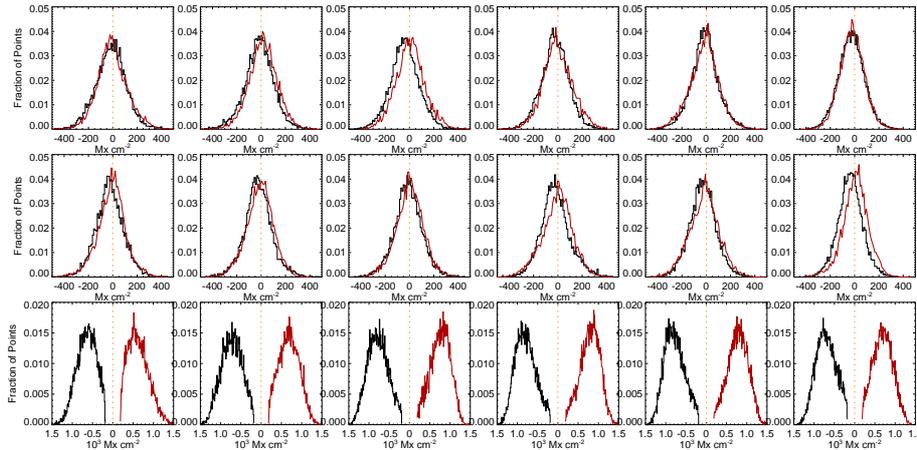}
\caption{Histograms of plage-identified areas of NOAA\,AR\,12457 for the {\it Hinode}/SP
Level-2 pipeline
data output, prepared as described in the text.  Left to Right: results from the six
scans (Table\,\ref{tbl:data}); central meridian crossing is at approximately 2015.11.25 00:00, 
between the $3^{\rm rd}$ and $4^{\rm th}$ scan. Red/Black indicate pixels 
that have positive/negative $\Bzh$; histograms are for $\Bxh$ (top), $\Byh$ (middle), and $\Bzh$ (bottom).
While there is some variation, the distributions for $\Bxh$ and $\Byh$ are essentially 
independent of polarity and location on the disk.}
\label{fig:ar12457_merlin_histos}
\end{figure}

\begin{figure}[h!]
\includegraphics[width=1.0\textwidth,clip, trim = 0mm 0mm 0mm 0mm, angle=0]{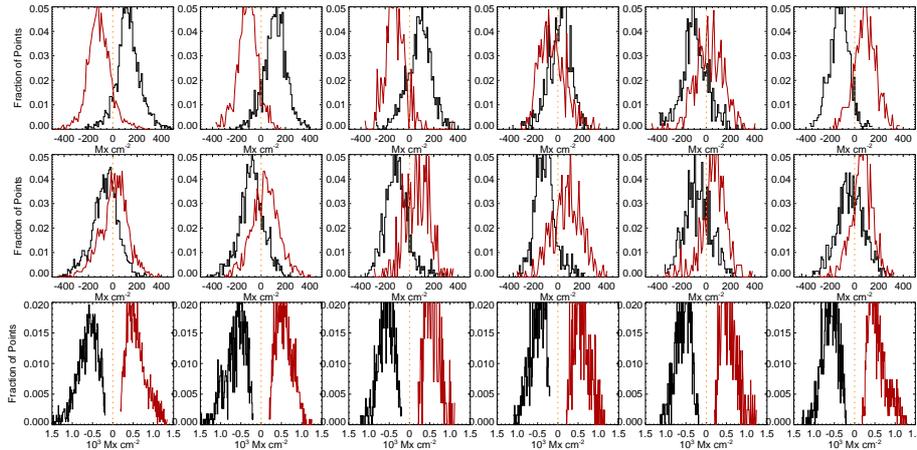}
\caption{Same presentation as in Figure~\ref{fig:ar12457_merlin_histos} and for the the same
field of view, but from {\it SDO}/HMI pipeline data, prepared as described in the text.  
Note the switch of the $\Bxh$ histogram peaks as the 
regions transits the central meridian, and the sustained
but consistent difference in $\Byh$.}
\label{fig:ar12457_hmi_histos}
\end{figure}

Similar to Figure~4 in \citet{btrans_bias_1}, histograms of the 
heliographic $\Bxh$ are presented separately for pixels with $\pm\Bzh$.
Without the presence of filament channels, nearby large active regions
or other phenomena that could introduce a physical direction preference,
the assumption is that the histograms should overlap: the distribution of $\Bxh$
or $\Byh$ should not differ according to either the sign of $\Bzh$ or viewing angle.
Tracking the plage in NOAA~AR\,12457 over 6 days, we see the very different behavior
between the {\it Hinode}/SP pipeline output (Figure\,\ref{fig:ar12457_merlin_histos})
and the {\it SDO}/HMI pipeline output (Figure\,\ref{fig:ar12457_hmi_histos}),
with consistent $<\!\Bxh\!>$ and $<\!\Byh\!>$ regardless of polarity and viewing
angle for the former, and a switch of $<\!\Bxh\!>$ between east/west hemisphere for the
latter, as well as a consistent offset in $<\!\Byh\!>$.


In situations such as NOAA~AR\,12457 where there is a sufficient sample of plage
with both polarities, the behavior is summarized by metrics describing the 
differences in the distributions of target magnetic components.  Specifically, 
metrics should target the behavior of the $\Bxh$ and of $\Byh$ most notably, plus
$\Bzh$ when it provides information.  The absolute separation $\overline{|\rm sep|}$
is the absolute difference of the medians of two distributions of the target $\Bxh, \Byh, or \Bzh$,
the two distributions being whether the underlying field has $\Bzh\!>\!0$ or $\Bzh\!<\!0$;
the mean absolute separation is taken over the 6 samples.  This is accompanied by the standard
deviation of the {\it signed} separation, $\sigma({\rm sep})$ which is larger for a 
switch in sign and smaller if, for example, the distributions are separated but do not 
change significantly between samples.  We additionally consider the median absolute
deviation (MAD; ${\rm med}(|\hat{X} - \tilde{X}|)$, where $\hat{X}$ is the median (most
probable) value for the target distribution) and $\tilde{X}$ is its median over the 6 
samples, and the maximum absolute deviation MaxAD which is the maximum difference
between the median of the target distribution (separately for $\Bzh\!>\!0$ or $\Bzh\!<\!0$)
from its expected value: 0.0 for $\Bxh$ and $\Byh$ and the mean $\Bzh$ over the
6 samples, as we hypothesize little evolution.

\begin{figure}
\centerline{
\includegraphics[width=0.65\textwidth,clip, trim = 0mm 0mm 13mm 0mm, angle=0]{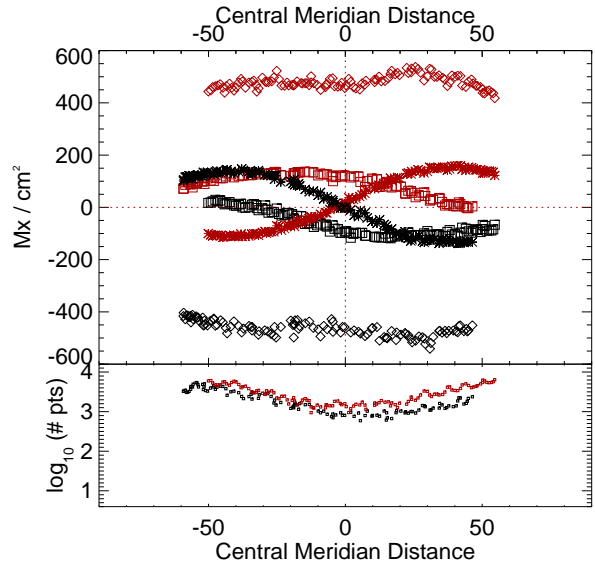}}
\caption{Timeseries of means of the distributions of $\Bxh (*), \Byh (\Box), \Bzh (\Diamond)$ 
for plage-identified concentrations in ``Plage 3'' (see Fig.\,\ref{fig:hmi_plage3}) and the 
standard {\it SDO}/HMI pipeline (``PIPE\_720s'') output, plus the number of points in the distributions (subplot).   
Red/black indicate positive/negative polarity $\Bzh$.}
\label{fig:vpipe_ts_plage3}
\end{figure}

\begin{figure}
\centerline{
\includegraphics[width=0.52\textwidth,clip, trim = 0mm 0mm 0mm 0mm, angle=0]{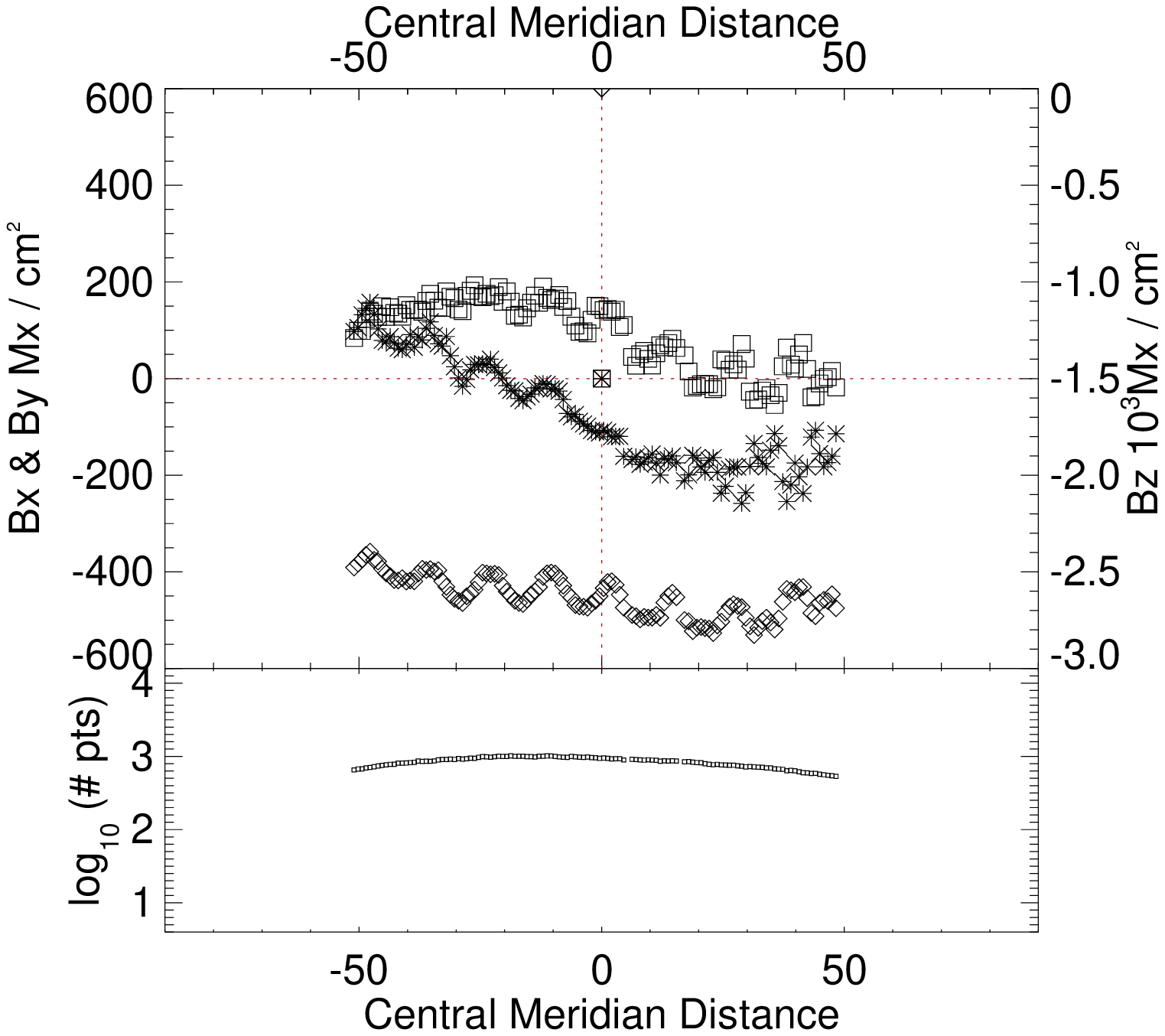}
\includegraphics[width=0.48\textwidth,clip, trim = 0mm 0mm 10mm 0mm, angle=0]{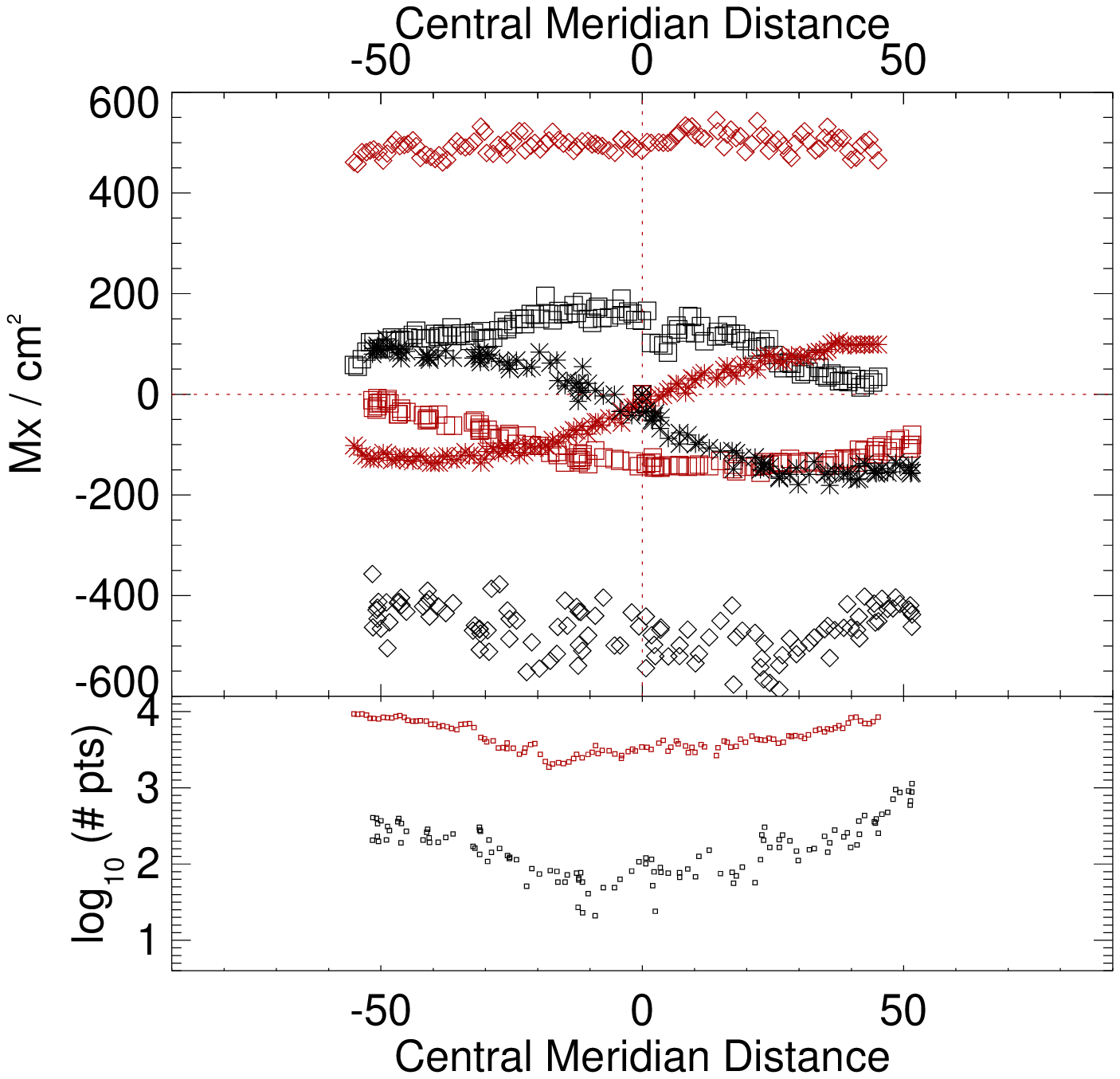}}
\centerline{
\includegraphics[width=0.52\textwidth,clip, trim = 0mm 0mm 0mm 0mm, angle=0]{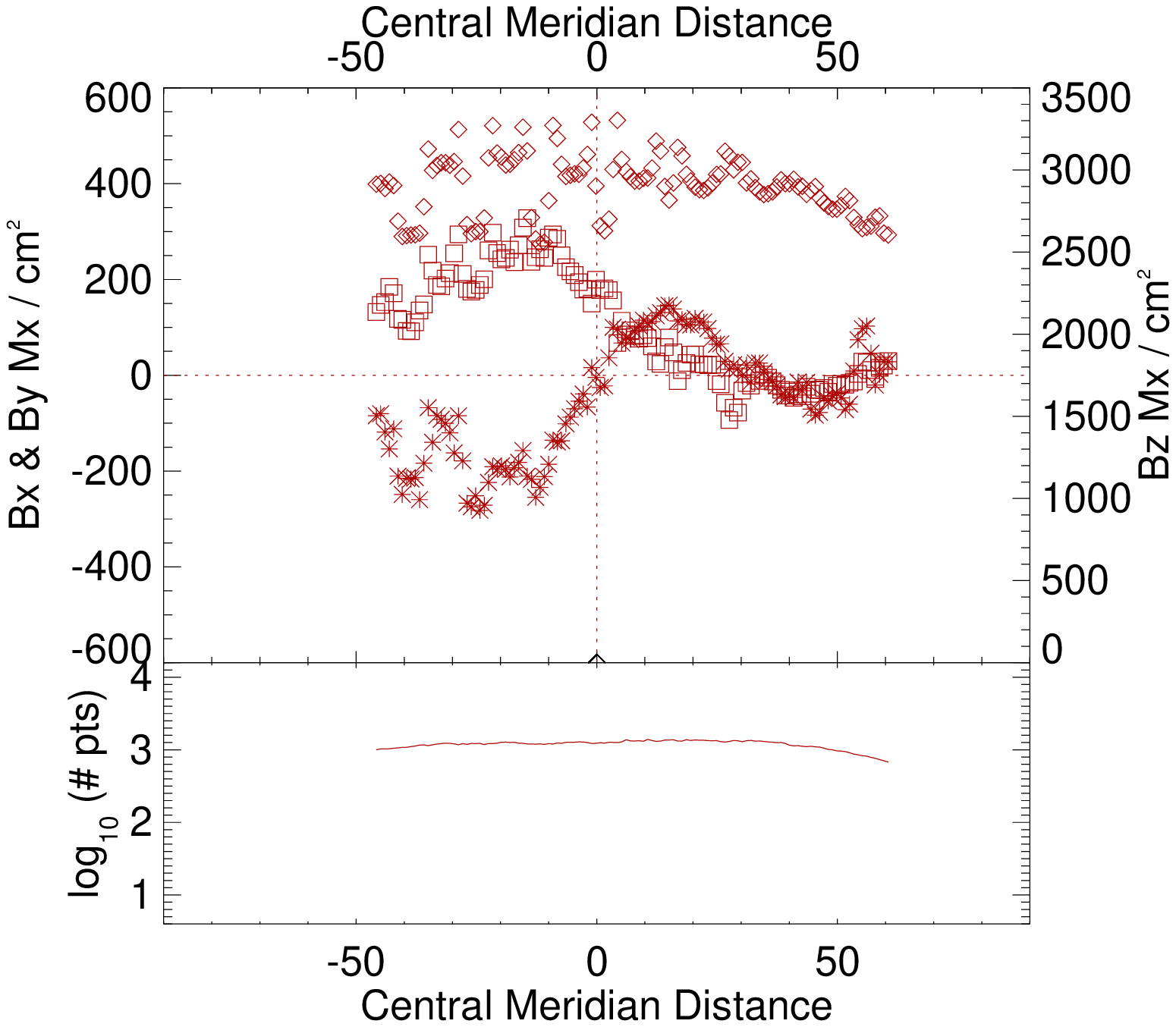}
\includegraphics[width=0.48\textwidth,clip, trim = 0mm 0mm 10mm 0mm, angle=0]{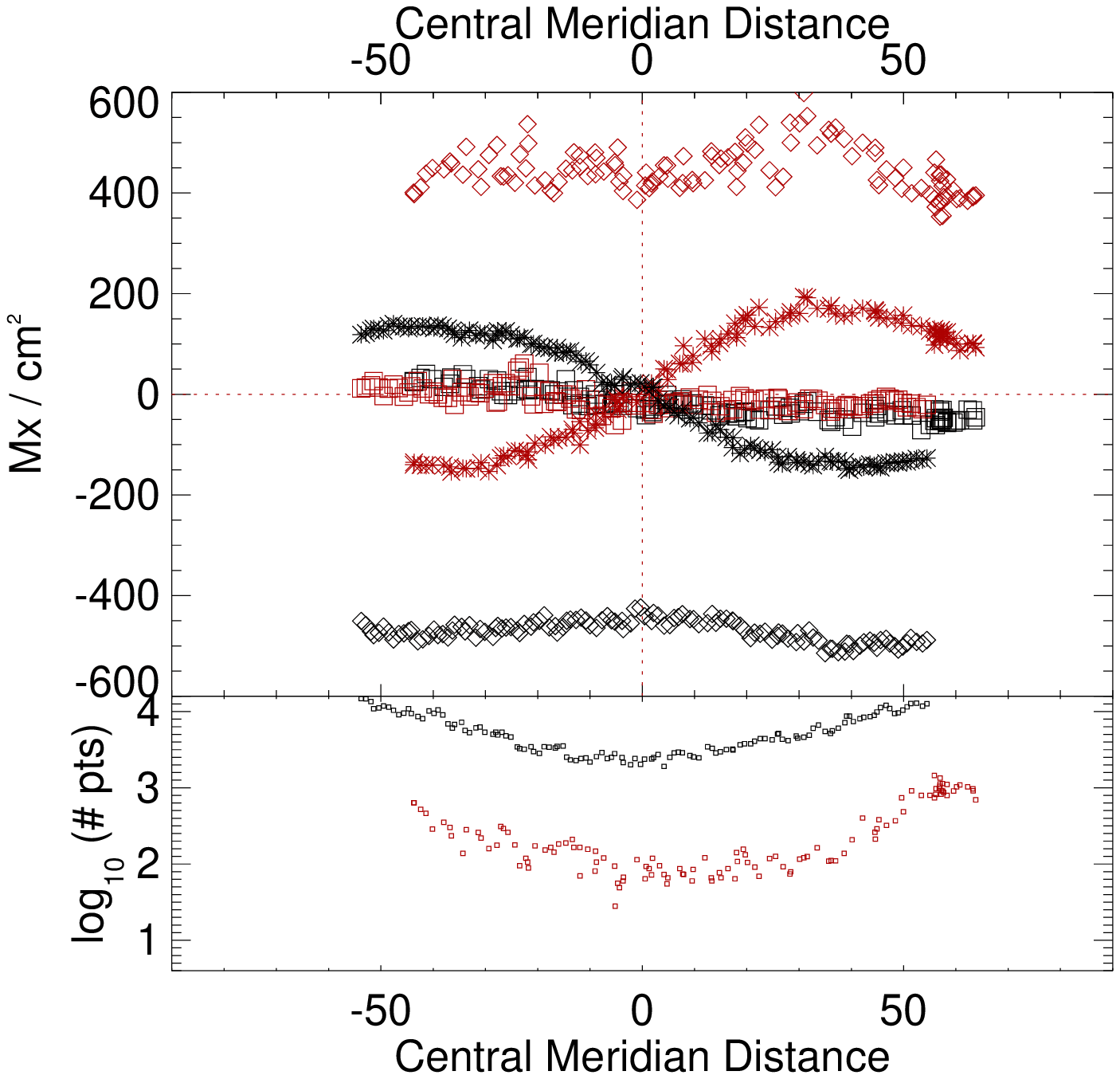}}
\centerline{
\includegraphics[width=0.52\textwidth,clip, trim = 0mm 0mm 0mm 0mm, angle=0]{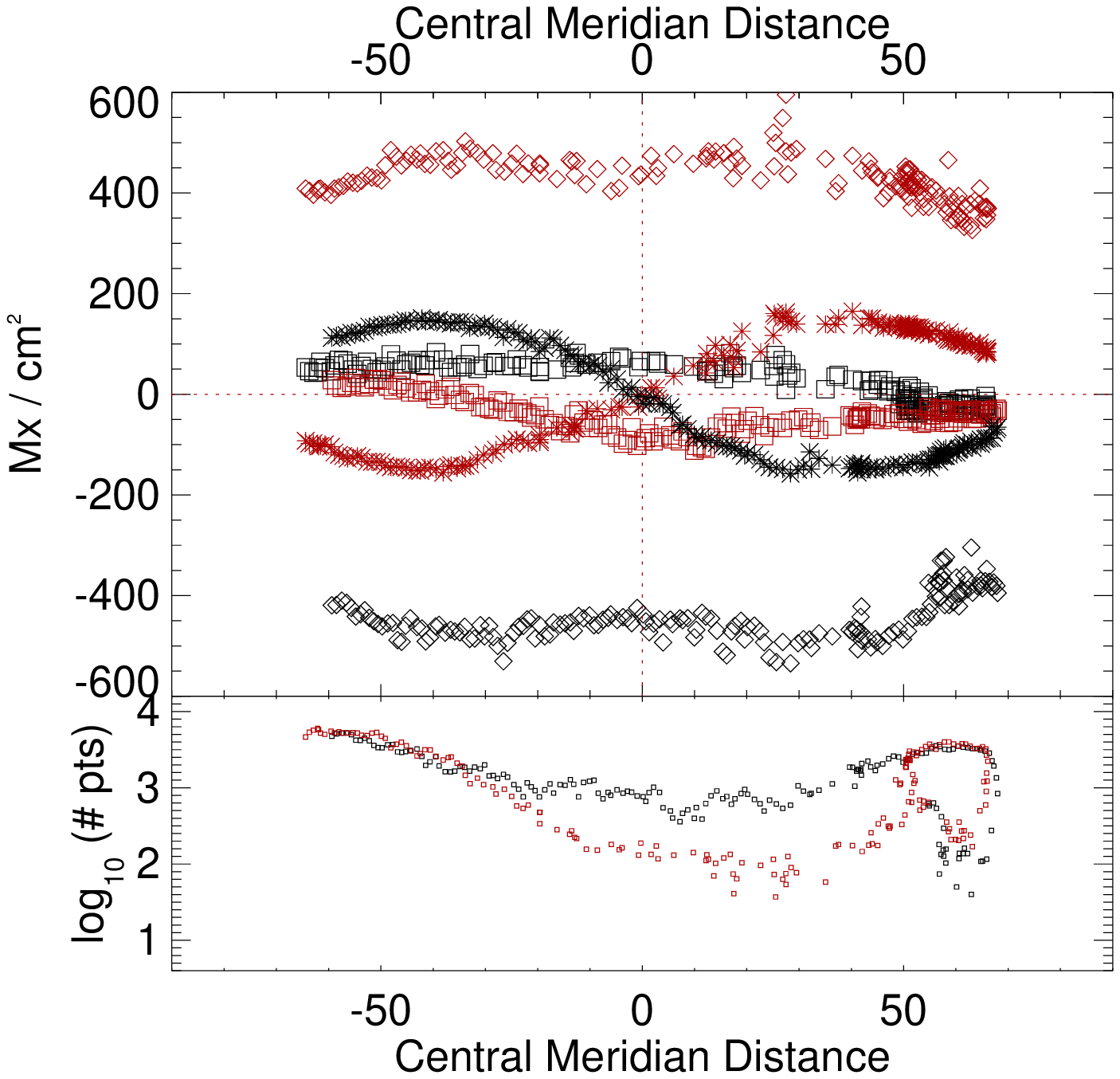}
\includegraphics[width=0.48\textwidth,clip, trim = 0mm 0mm 10mm 0mm, angle=0]{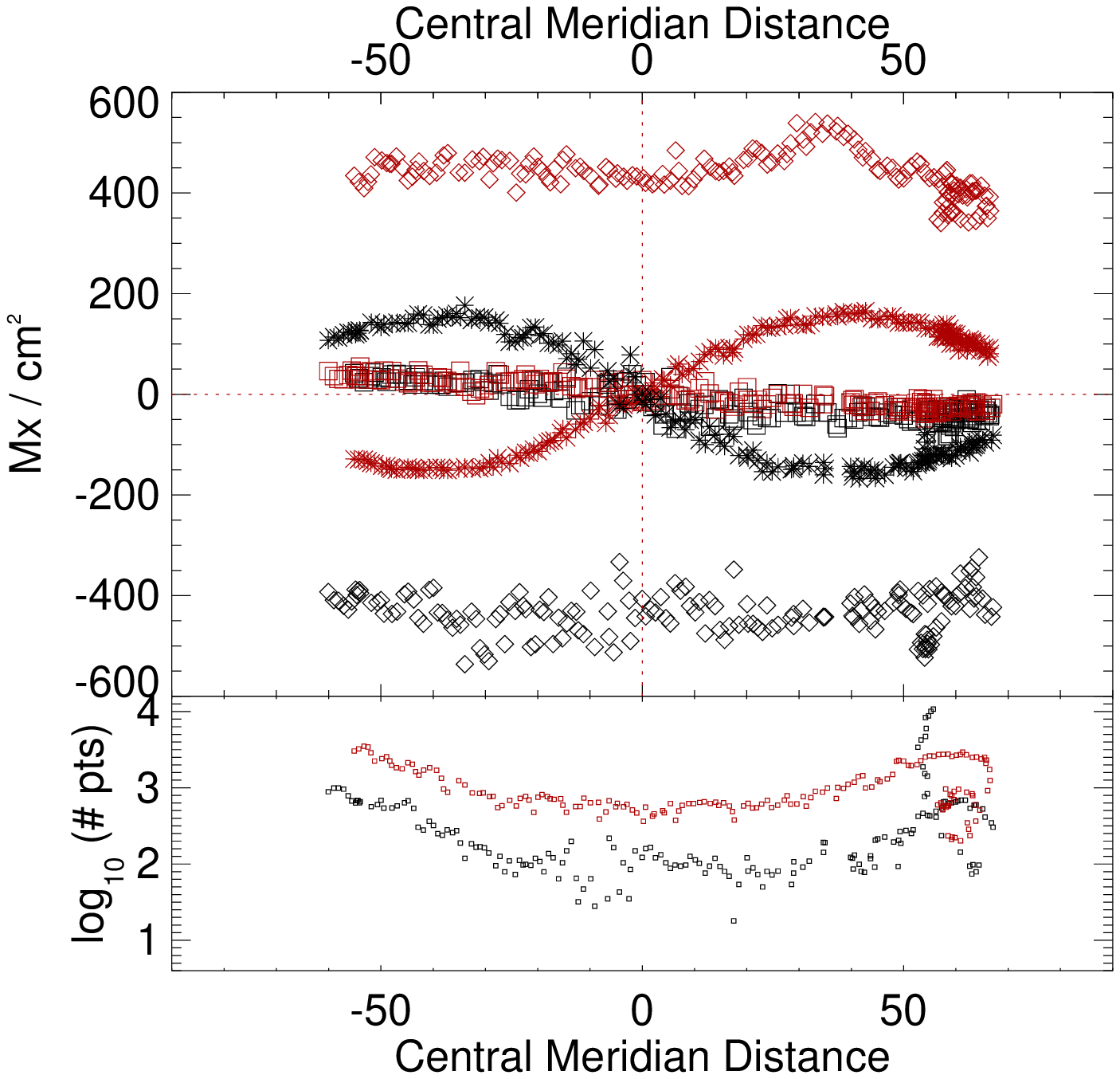}}
\caption{Same presentation as Fig.\,\ref{fig:vpipe_ts_plage3}: $\Bxh (*), \Byh (\Box), \Bzh (\Diamond)$, 
Red/black indicate positive/negative polarity $\Bzh$. (Top, left/right): Spot 1, Plage 1, 
(Middle, left/right): Spot 2, Plage 2, (Bottom, left/right) Plage 4 North, Plage 4 South.
For the spot-related plots, note the different scale (right-hand y-axis) for the $\Bzh$ component.}
\label{fig:vpipe_ts_rest}
\end{figure}

This analysis is also applied to the sub-area targeted patches of
{\it SDO}/HMI time-series data.  In Figures\,\ref{fig:vpipe_ts_plage3} and \ref{fig:vpipe_ts_rest}
we summarize the behavior of the target distributions as a function of the mid-point longitude
for all three vector components.   Quantitative summaries take the same form as for the
NOAA\,AR\,12457 analysis: mean absolute separation, standard deviation of the signed separation,
MAD, and MaxAD.

\subsubsection{Distributions of $|\gamma^i|$ with Observing Angle}
\label{sec:incl_distrib_metric}

Ideally there should be metrics available from the 
image-plane components themselves, without requiring the disambiguation step.
If we describe the bias as an over- (or under-) estimate in the $\Bperp$
component, then this will manifest in the image-plane inclination angle
$\gamma^i$.  It is in this context that focusing on plage and the
radially-oriented locations
within stable sunspots becomes useful, because we can assume that
for radially-directed fields, $\Br = \Bpar / \mu$, where $\mu = \cos(\theta)$ 
the cosine of the observing angle \citep{Svalgaard_etal_1978,WangSheeley1992}.
In other words, for radially-directed fields without bias, 
$\gamma^i = \mu$.  

\begin{figure}
\includegraphics[width=1.0\textwidth,clip, trim = 0mm 0mm 2mm 2mm, angle=0]{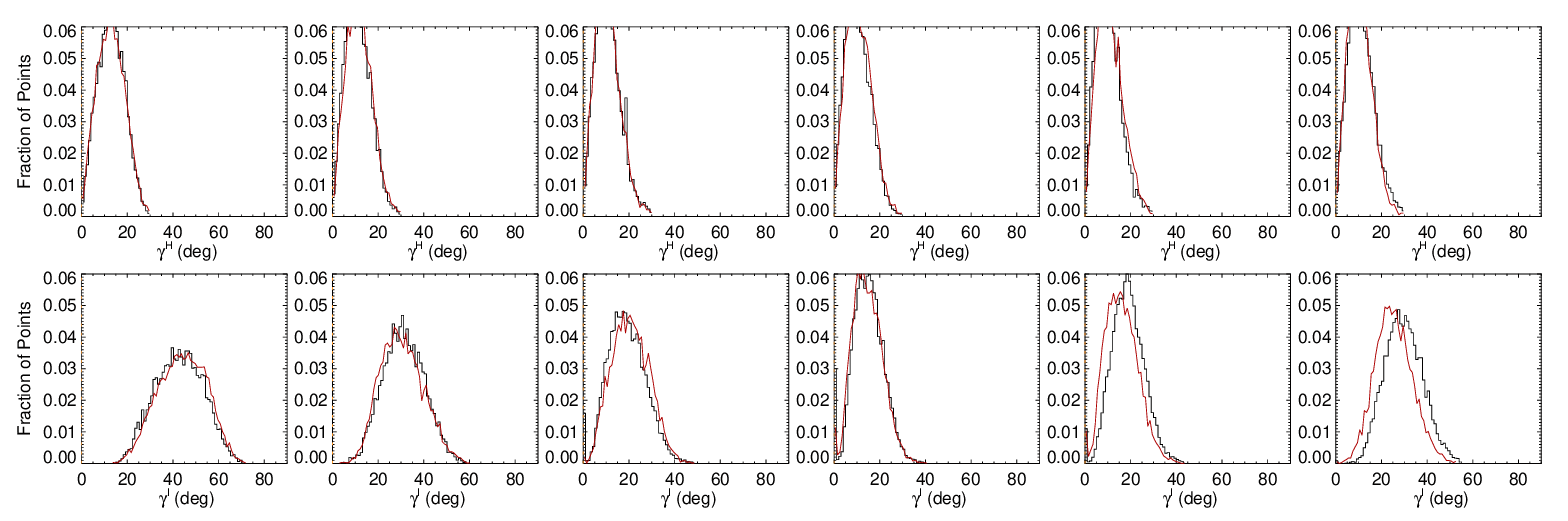}
\caption{Same presentation as in Figure~\ref{fig:ar12457_merlin_histos} for {\it Hinode}/SP \merlin\ 
pipeline output data, but showing the distribution of $\gamma^h$ (top) and $\gamma^i$ (bottom).} 
\label{fig:ar12457_merlin_incl_histos}
\end{figure}

\begin{figure}
\includegraphics[width=1.0\textwidth,clip,trim=0mm 0mm 2mm 2mm,angle=0]{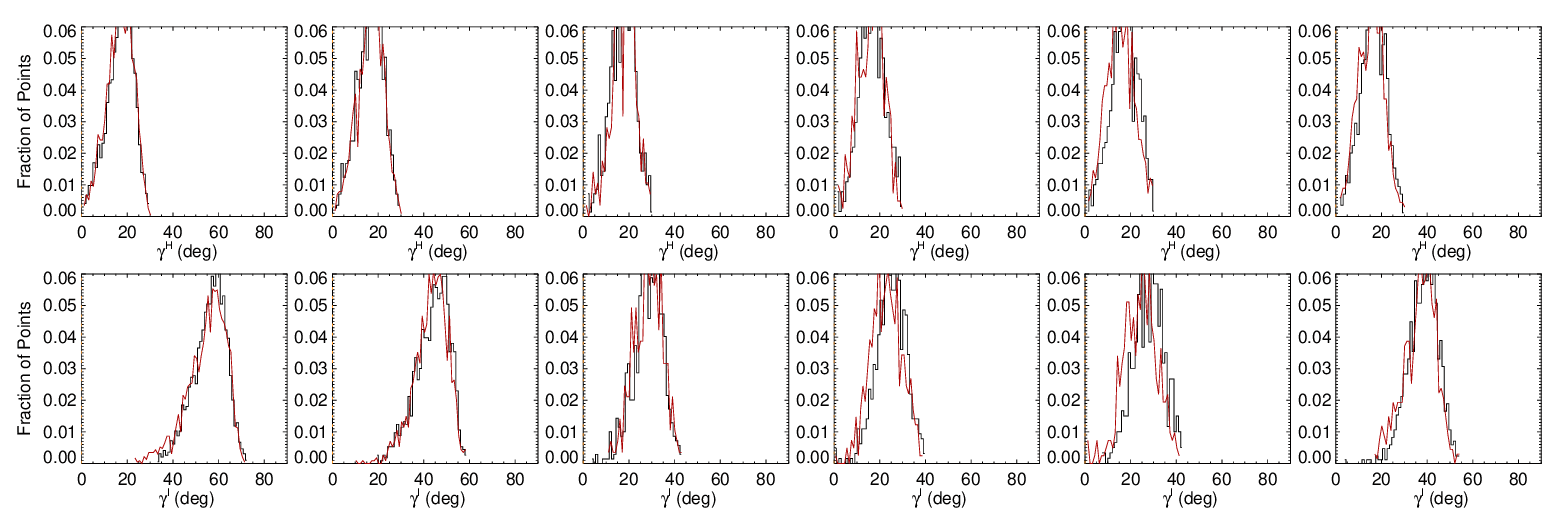}
\caption{Same general presentation as Figure~\ref{fig:ar12457_hmi_histos} for 
{\it SDO}/HMI pipeline \vfisv\ output data, but showing the distribution of $\gamma^h$ (top) 
and $\gamma^i$ (bottom) as per Figure~\ref{fig:ar12457_merlin_incl_histos}.}
\label{fig:ar12457_hmi_incl_histos}
\end{figure}

First we examine the heliographic (local, or physical) inclination
angle of the target plage points in the NOAA\,AR\,12457 data
and confirm that the distributions are the same between the
positive and negative regions, change minimally as a function
of east/west location, and while not exactly radial, are not
particularly inclined (Figures\,\ref{fig:ar12457_merlin_incl_histos},
\ref{fig:ar12457_hmi_incl_histos}, top panels).

The distribution of $|\gamma^i|$, however, distinctly
changes shape as a function of observing angle, and
differs between {\it e.g.} the {\it SDO}/HMI and {\it Hinode}/SP pipeline output
(Figures~\ref{fig:ar12457_merlin_incl_histos},\,\ref{fig:ar12457_hmi_incl_histos},
lower panels).  The skew of the distributions shows exactly opposite
behavior with east/west location between the two pipeline data outputs.

Figures\,\ref{fig:ar12457_merlin_incl_histos},
\ref{fig:ar12457_hmi_incl_histos} (top panels)
confirm that there is a distribution of orientations
within these structures, of course, but they 
are dominated by radially-directed field.
Hence we focus on $E(\gamma^i)$, the expected value or 
most probable value, and whether it tracks the
observing angle $\mu =\cos(\theta)$ (Figure~\ref{fig:gini_eg}).
The degree to which $E(\gamma^i)$ points {\it vs.} $\theta$ lie off the $x=y$ 
line provides information on the bias, including whether there is an under- or over-
estimation of $\Bperp$.  This key diagnostic can be performed on
limited-FOV time-series data for (statistically) unchanging structures,
but can also be used for full-disk data without requiring any time-series,
provided the required structures are present.

In Figure~\ref{fig:gini_eg}, we see that the \merlin\ $E(\gamma^i)\ {\rm vs.}\ \theta$ 
results align with 
$x=y$ nearly perfectly, while the three examples from the {\it SDO}/HMI pipeline 
all show significant deviations, although with inconsistent functional form.
The differences between the two plage-targeted examples could be due to
different observing epochs (2010 {\it vs.} 2016), or subtle changes in the
tracked structures due to evolution or field-of-view.  The 
deviation from $x=y$ for the sunspot confirms the presence of the bias
in structures that have strong, pixel-filled signal, presumably with $f\!f=1.0$.
The different form of $E(\gamma^i)\ {\rm vs.}\ \theta$ for the spot target
compared to plage targets may be due to a combination of signal/noise ratio,
signal saturation, scattered light treatment, plus 
the different impact of bias according to the treatment of $f\!f$.

The primary metric then to evaluate $|\gamma^i|$ distributions for
plage as a function of $\mu=\cos(\theta)$ is the Gini coefficient $\mathcal{G}$ or Receiver (Relative)
Operating Characteristic (ROC) Skill Score \citep[ROCSS;][]{JolliffeStephenson2012,nci_daffs}
$\mathcal{G}=2*{\rm AUC} -1$, where the AUC metric is the integrated area under the curve.
Provided sufficient coverage in $\mu$ is available, $\mathcal{G}$ essentially
measures a signed area departure from the $x=y$ line.  Other metrics
such as the the maximum absolute deviation could also be used
here, but $\mathcal{G}$ is intuitive, for $\mathcal{G} > 0$ indicates
over-estimation of $\Bperp$ whereas $\mathcal{G} < 0$ indicates an
underestimation.  To compute $\mathcal{G}$, we normalize the ranges to $[0,1]$
and extend the sampled ranges to those ends for all data, in order to
cursorily account for different sampling in $\mu=\cos(\theta)$ between
datasets and tests.  The choice of using angles in degrees but normalized
by 90 simply provides an intuitive presentation and AUC calculation; the
same information is available using $\mu$ and $\cos(|\tilde{\gamma^i}|)$,
similarly both normalized to $[0,1]$.

\section{Results}
\label{sec:results}

There are a large number of tests that were performed; here we summarize
results for targeted questions, first the experiments with model data,
then observational.

\subsection{The Magnitude and Impact of the Problem}

The most obvious impact of this bias manifests in the sign-flip
of $\Bxh$ as extended patches of unresolved structures rotate
east/west across the solar disk, as described here and in previous
papers \citep{btrans_bias_1,Liu_etal_2022}.    The magnitude is
demonstrated here quantitatively (Figures~\ref{fig:ar12457_hmi_histos},
\ref{fig:vpipe_ts_plage3}, \ref{fig:vpipe_ts_rest}):  up to a
${\rm few} \times 100\,{\rm  Mx\,cm}^{-2}$ signal of bias is present, 
or almost 30\% of the total pixel-averaged field magnitude for the plage regions.

It was surmised in those papers that there should be a bias in the
north/south distribution of $\Byh$ as well.  It is shown quantitatively
here (same figures).  The $\Byh$ component behaves in an opposite way
for plage points with $\Br > 0$ {\it vs.} $\Br < 0$ when located north
{\it vs.} south of the equator.  The effect is somewhat subtle due to
the limited north/south extent available for analysis, but it is clear.

\begin{figure}
\centerline{
\includegraphics[width=0.5\textwidth,clip, trim = 5mm 2mm 2mm 5mm, angle=0]{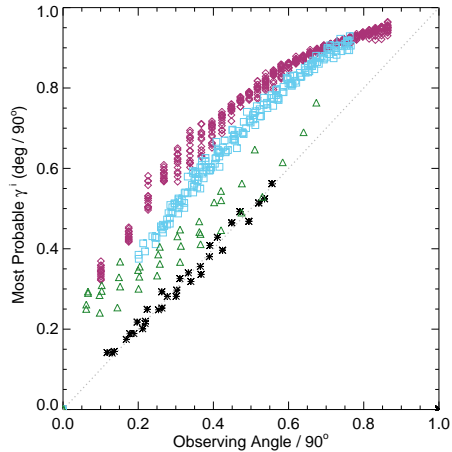}}
\caption{Plot of the most probable image-plane absolute inclination
$|\gamma^i|$ (y-axis) for plage areas in rings or bins of $\Delta\mu
= 0.025$ as a function of the central value for that ring (x-axis);
both axes are in degrees but normalized by 90$^\circ$.  Shown are {\it Hinode}/SP 
\merlin\ Level-2.0 data for NOAA\,AR\,12457 for all days sampled (black $\ast$), 
and three targets from the {\it SDO}/HMI \vfisv\ pipeline:  full-disk data for 
2016.05.18 (red $\Diamond$), ``Plage \#3'' (blue $\Box$), and ``Spot \#2'' 
(green $\bigtriangleup$).
Dashed line: $x=y$.  The {\it SDO}/HMI full-disk data are sampled for 24hr at 
96\,m cadence (15 samples) and show as well the {\tt OBS\_VR}-related variation
\citep{hmi_pipe,Schuck_etal_2016}.}
\label{fig:gini_eg}
\end{figure}

It was also briefly mentioned in previous papers that there may be an impact
on the inferred radial component $\Bzh$ or $\Br$  (the component most
widely used for global modeling presently), and this is also demonstrated
here.  It is a smaller effect still, but can be seen as subtle peaks in the
most probable pixel-averaged flux away from disk center, and a dip in
the magnitude near disk center.

In summary, because this is a problem in image plane, it impacts all
vector components in physical space, including $\Br$ ($\Bzh$).  It will
also lead to incorrect or biased estimates of any derived quantity
such as the vertical current, the force-free parameter $\alpha$, {\it etc.}.
The bias is strongest in {\it unresolved} structures -- and as such
will constitute significant portions of active regions as well as the 
large-scale plage regions.

It was stated in the earlier papers that there was no bias in the
strong-field or $f\!f \approx 1.0$ areas.  We find this is not actually the
case.  We show that in the central portions of two large, round, stable
sunspots, there is a non-zero most probable inclination.  The direction
and magnitude of the most-probable magnetic field vector changes sign (or
nearly changes sign) for $\Bxh$ as the spots transit the central meridian,
and displays a non-zero difference in the most probable value of $\Byh$
as well.  The variations in $\Bzh$ are smaller than the orbital-velocity
related variations and cannot be confirmed here.

The bias in $f\!f \approx 1.0$ areas may be caused ultimately by unresolved field
or the detailed handling of scattered instrumental light \citep{ivm3}.
It indeed fails to cause a full sign reversal of the components, that is true.
However we disagree with \citet{Liu_etal_2022} that ``This bias does not occur in strong 
field regions in sunspots.'' The signature of the bias is present, but only apparent 
with a quantitative examination of the field-component distributions. 

\subsection{Results from Model Data Experiments}
\label{sec:model_tests}

The simple toy model of plage-like distribution of field across a
range of observing angles is used to test three possible contributing
bias distributions: (1) a constant magnitude of bias added to 
$\Bperp$, (2) a contribution to $\Bperp$ that is a function of total
field, and (3) a contribution to $\Bperp$ that is a function of observing
angle, mimicking that which would be expected with systematically
higher fill factor derived in regions away from disk center (hence
the bias introduced by an imposed $f\!f=1.0$ will be greater
toward disk center).  The goal
here is to reproduce some of the quantitative characteristics observed:
the changing skew of $\gamma^i$ distributions with observing angle,
confirm the behavior of the Gini coefficient with bias and its sign, and
understand the source of the sometimes-observed curvature of $\gamma^i$
as a function of observing angle (Figure~\ref{fig:gini_eg}).  In most cases, 
both positive and negative bias is added.

\begin{figure}
\centerline{
\includegraphics[width=0.50\textwidth,clip, trim = 4mm 0mm 0mm 5mm, angle=0]{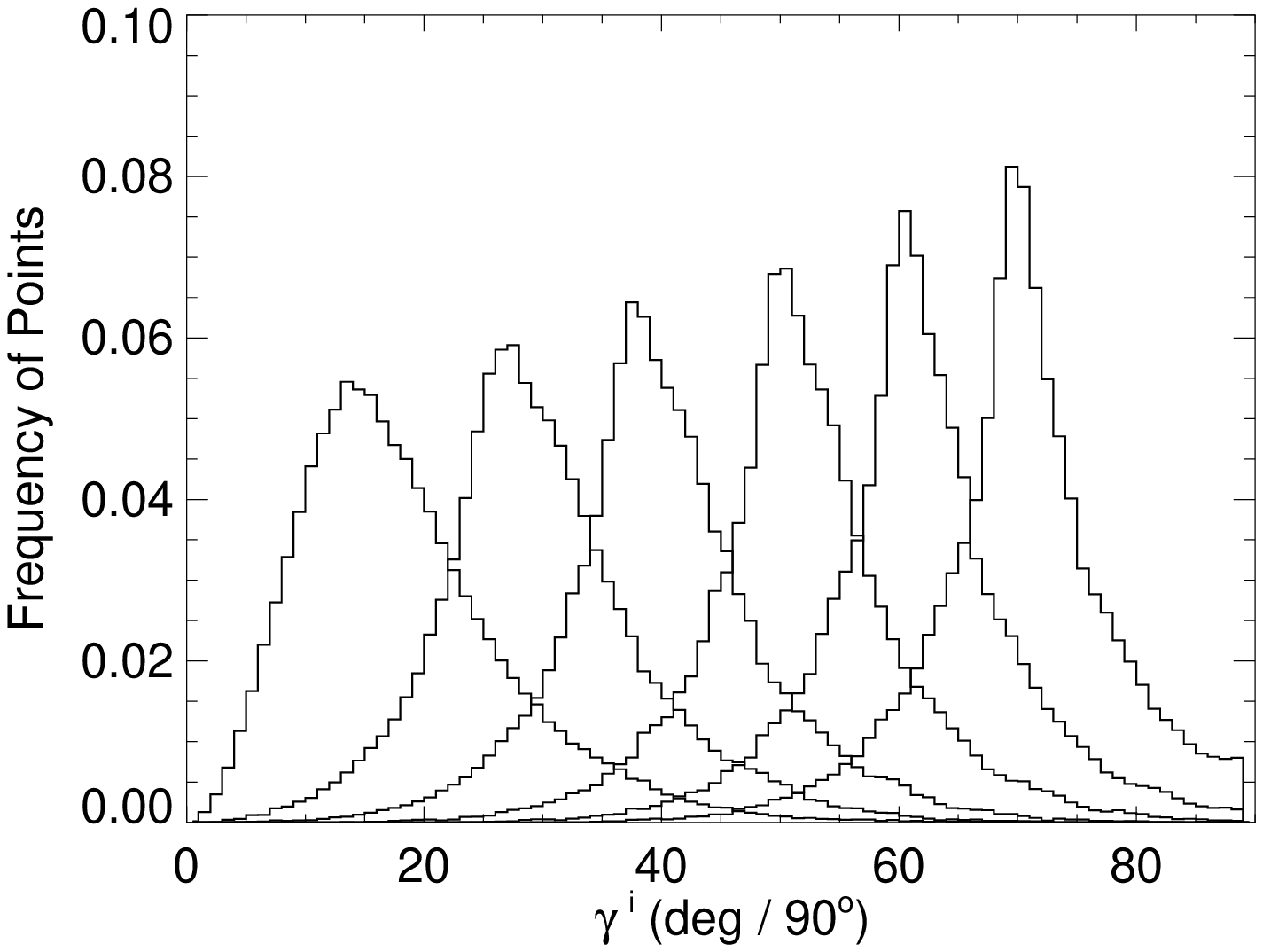}
\includegraphics[width=0.50\textwidth,clip, trim = 4mm 0mm 0mm 5mm, angle=0]{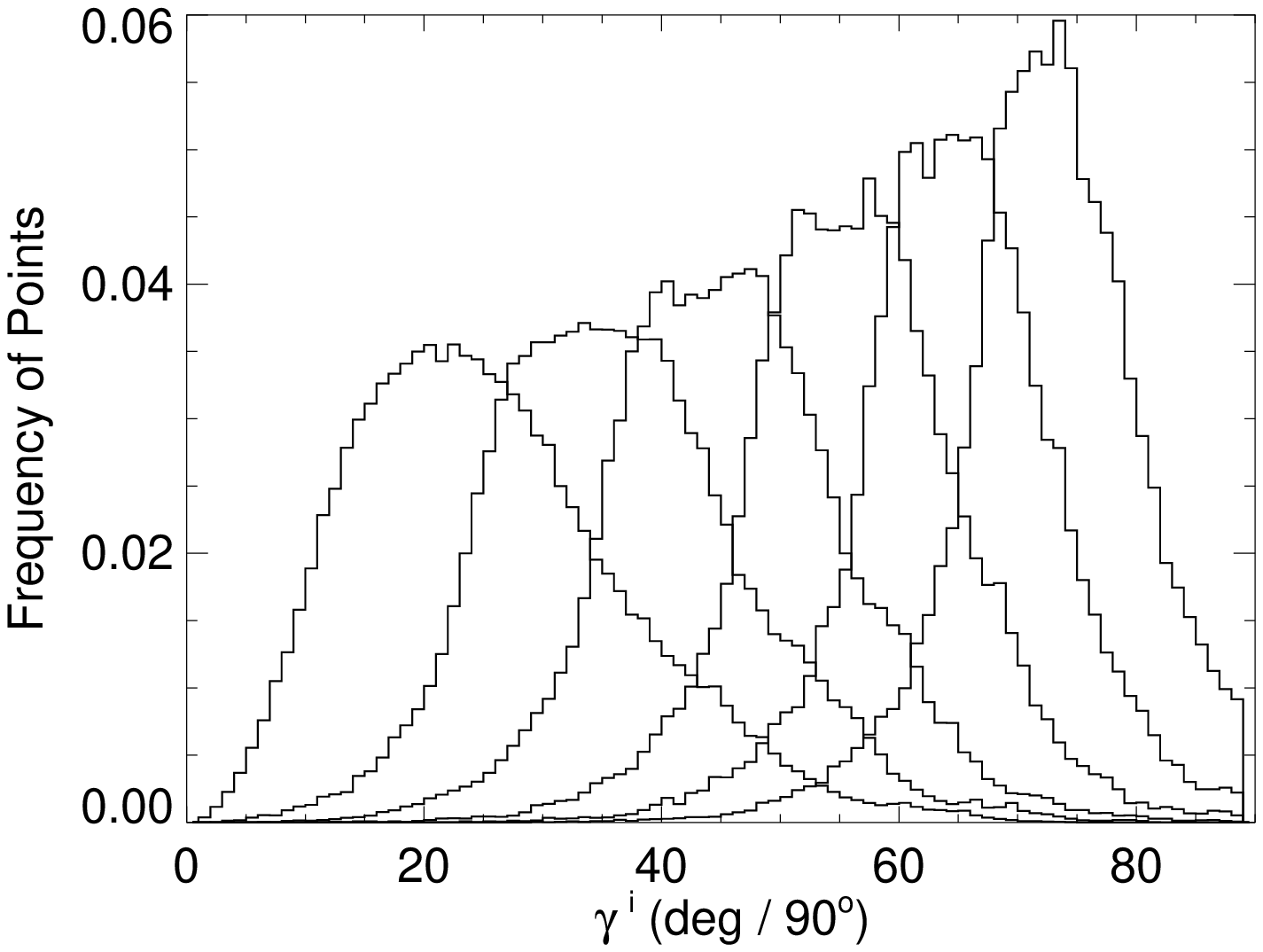}}
\centerline{
\includegraphics[width=0.50\textwidth,clip, trim = 4mm 0mm 0mm 5mm, angle=0]{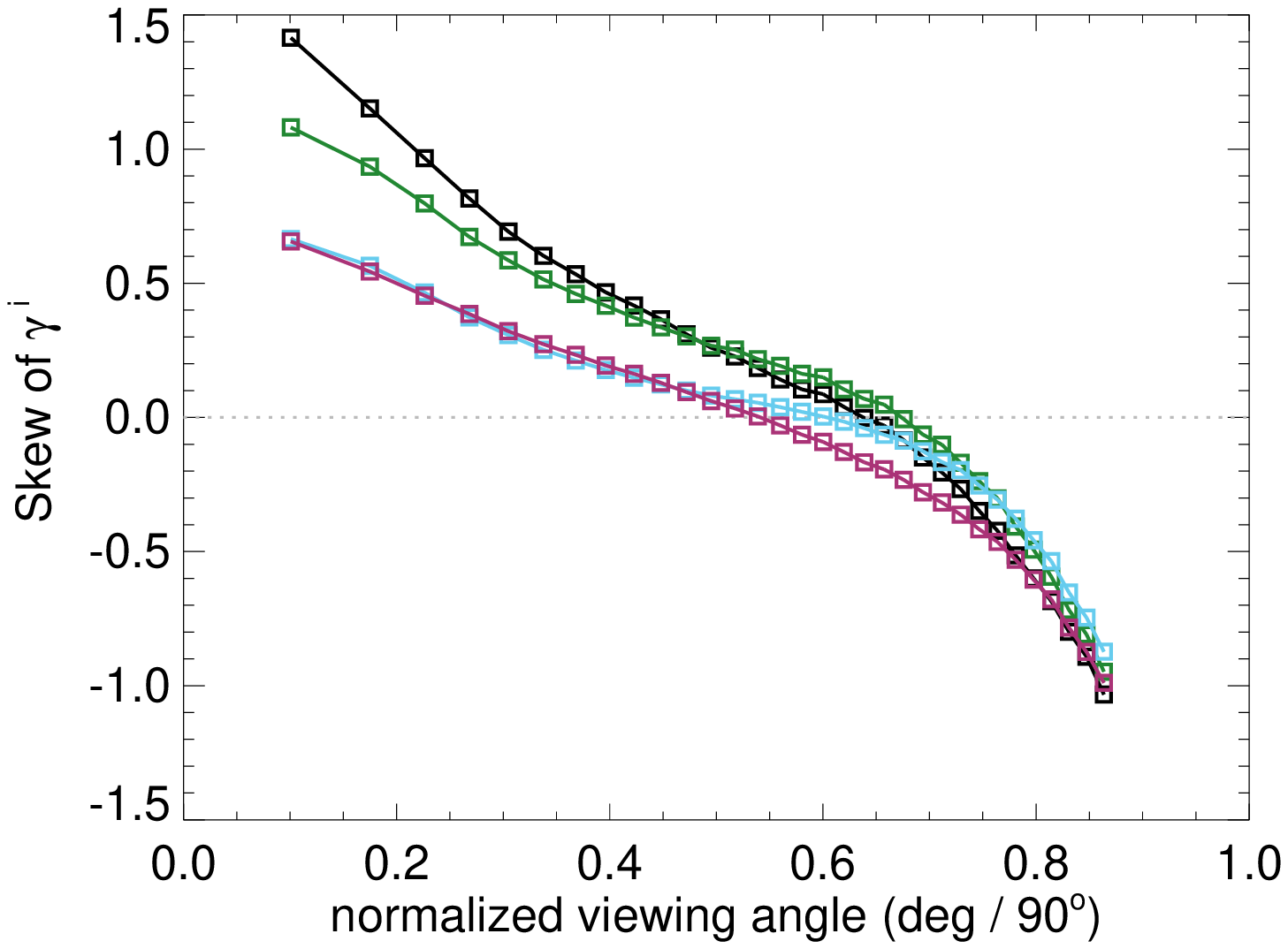}
\includegraphics[width=0.50\textwidth,clip, trim = 4mm 0mm 0mm 5mm, angle=0]{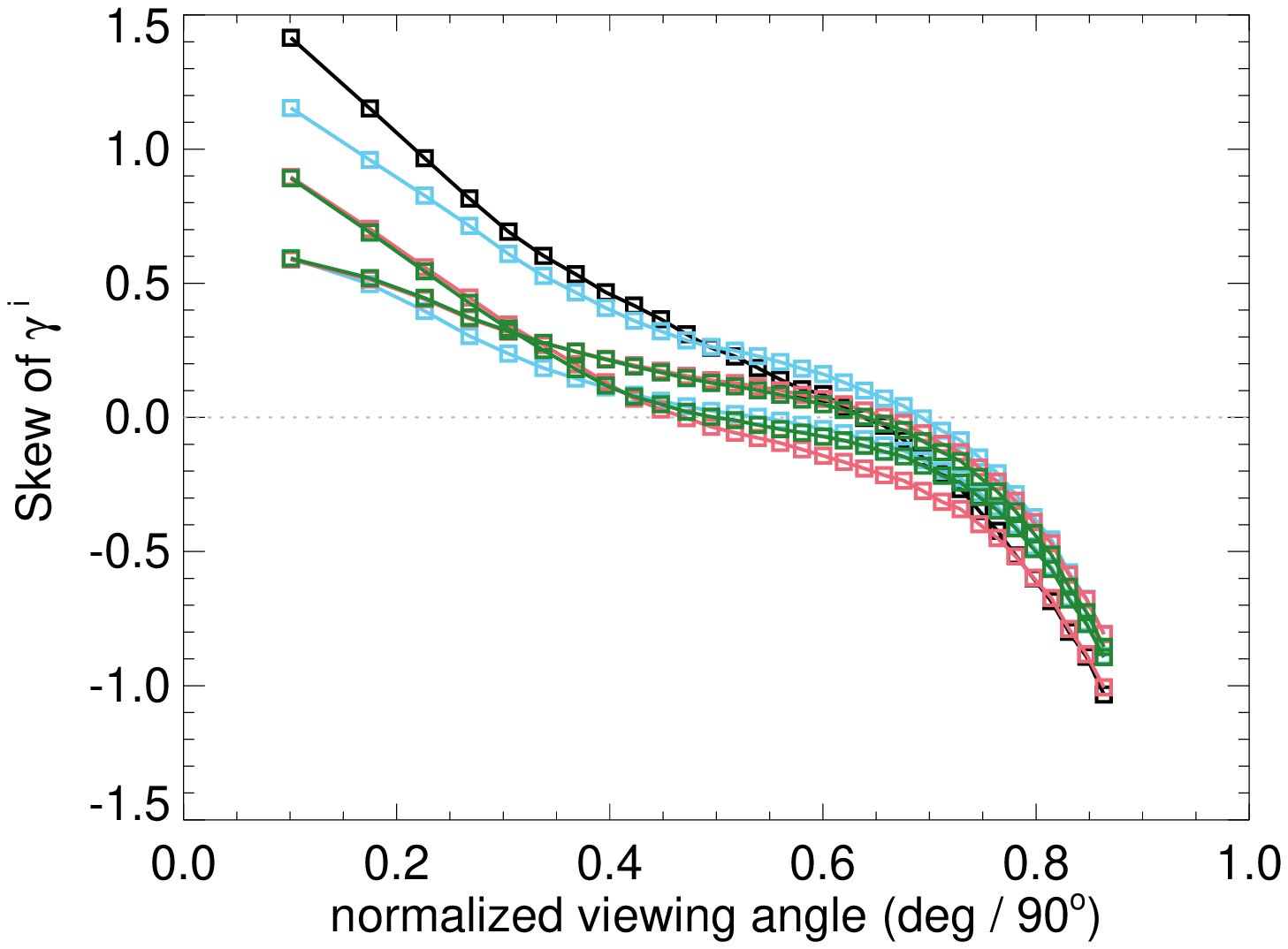}}
\caption{{\bf Top:} histograms of the ``observed'' $|\gamma^i|$ of the model data for
bins that are 0.025-wide centered at $\mu=[0.3625, 0.50125, 0.6625, 0.8125, 0.9125, 0.9875]$.
(Left): ``low-noise'' case, (Right): ``high noise'' case.
{\bf Bottom}: The skew of the $|\gamma^i|$ distributions as a function
of viewing angle as the result of different experiments.
(Left): effect of different levels of photon noise, no additional bias added.
Black Squares: no noise at all; also shown are ``low noise'' (green), 
``medium noise'' (light blue) and ``high noise'' (red).  
(Right): tests of different characterization of bias are shown, as related
to the same Black squares (no noise or bias).  Red: the experiment (1), a constant
magnitude of bias at $\pm 200$ added to $\Bperp$; Light Blue: (2), a bias that is $\pm25$\% of
the total field strength; Green: (3) a bias of $\pm 200 \times \mu$.}
\label{fig:model_skewstuff}
\end{figure}

The distributions of $|\gamma^i|$ with observing angle show a general
trend as is seen in the {\it Hinode}/SP data (see Figure\,\ref{fig:ar12457_merlin_incl_histos})
of a changing shape with observing angle (Figure\,\ref{fig:model_skewstuff}, Top).
The changes are indeed influenced by the degree of photon noise present.

Tracking this change in skew as a function of the experiments performed
regarding noise and bias is shown in Figure\,\ref{fig:model_skewstuff}, Bottom.
There is a distinct form of the change in skew of the $|\gamma^i|$ distribution
as a function of viewing angle, without any additional noise.  
The addition of photon noise and bias that originates with according to the three experiments
present departures from the no-noise case.  However, the `optimal' curve
(the no-noise no-bias case) is not straightforward to describe: it is not a simple
function of observing angle.  The curves with added noise and bias depart
from the optimal, but in sometimes subtle and non-unique ways.  As such,
we elect to not use the skew of the $|\gamma^i|$ distributions as 
a quantitative test of bias in the observational data.

\begin{figure}
\centerline{
\includegraphics[width=0.50\textwidth,clip, trim = 5mm 2mm 6mm 6mm, angle=0]{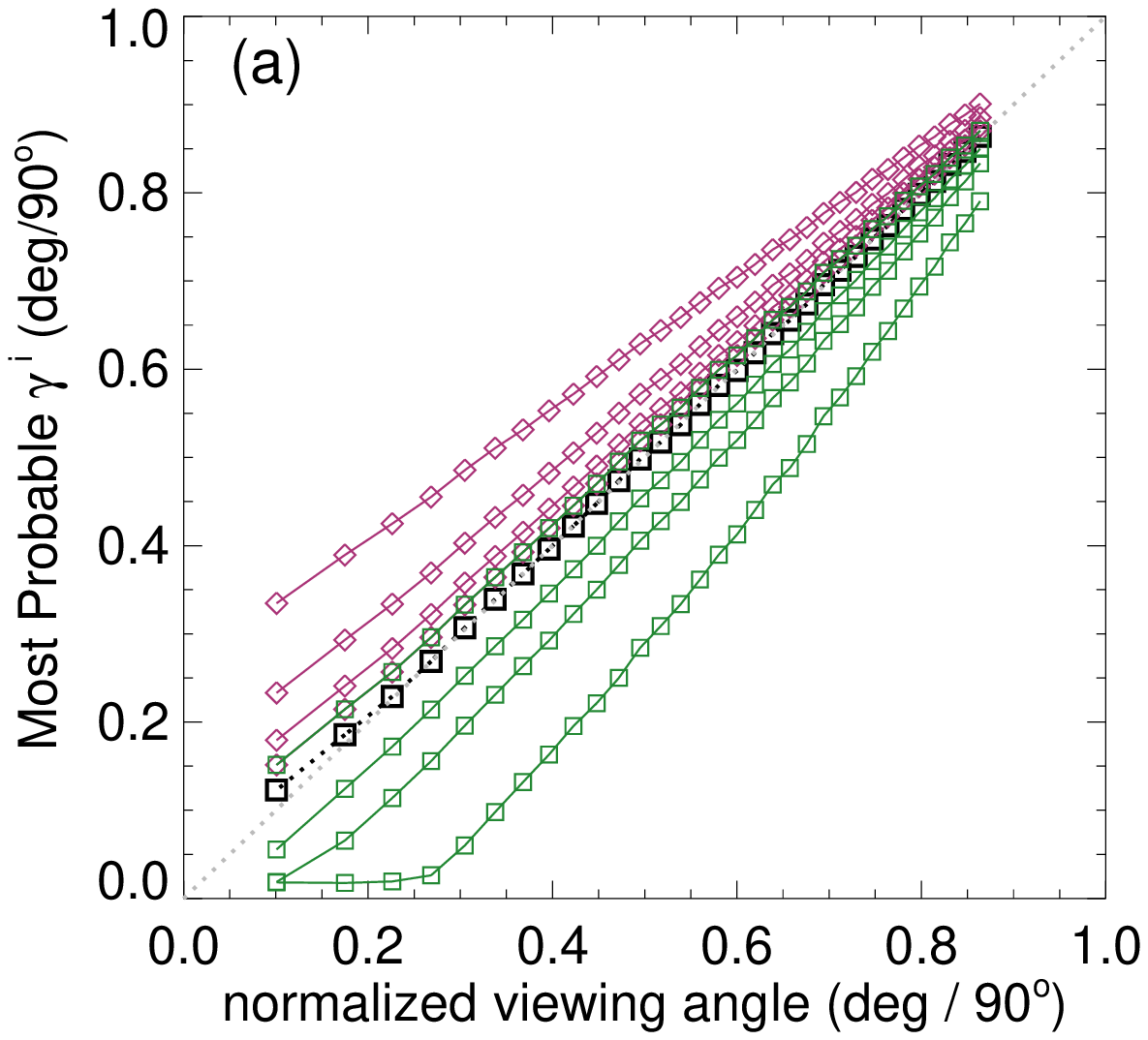}
\includegraphics[width=0.50\textwidth,clip, trim = 5mm 2mm 6mm 6mm, angle=0]{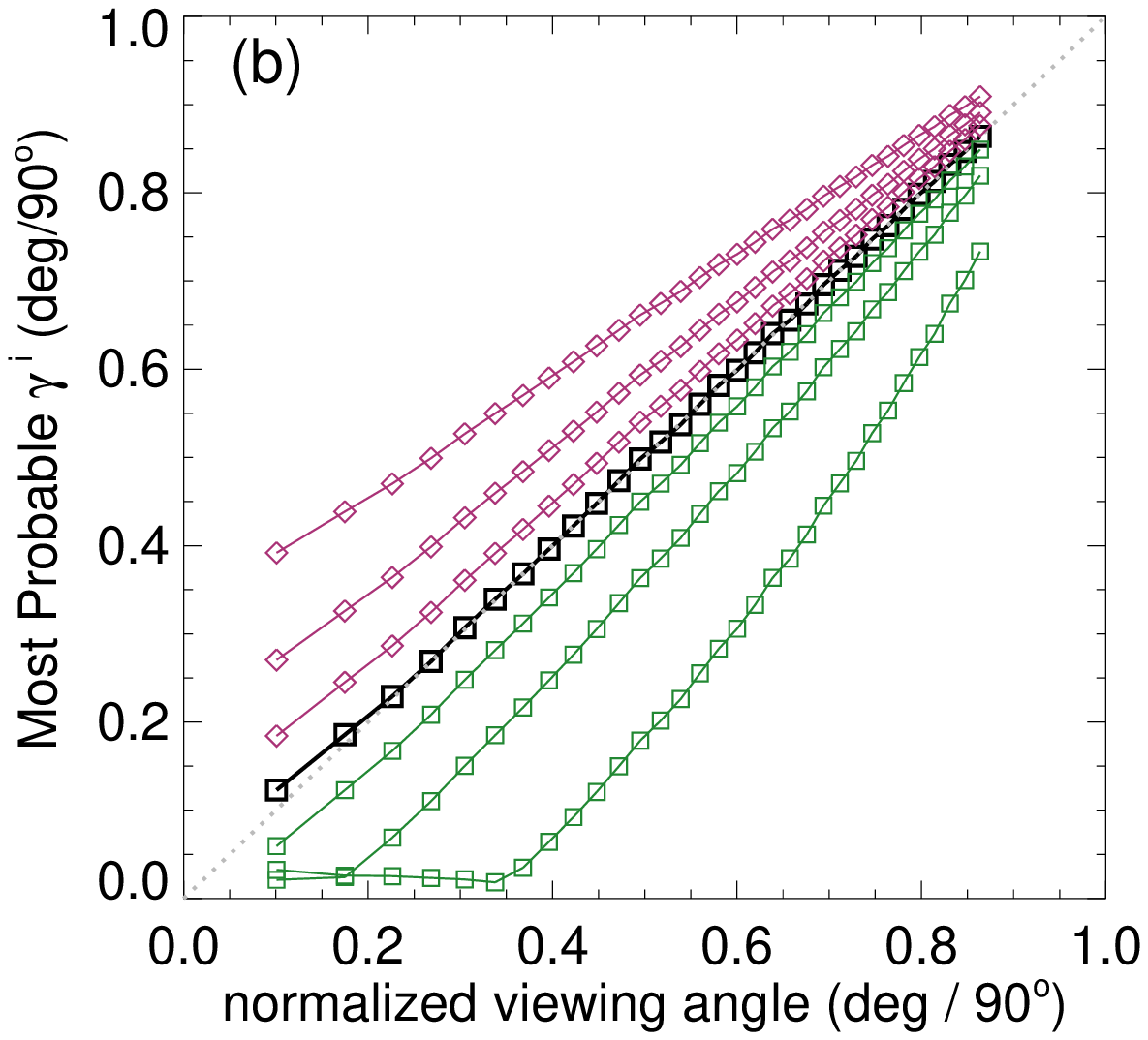}}
\centerline{
\includegraphics[width=0.50\textwidth,clip, trim = 5mm 2mm 6mm 6mm, angle=0]{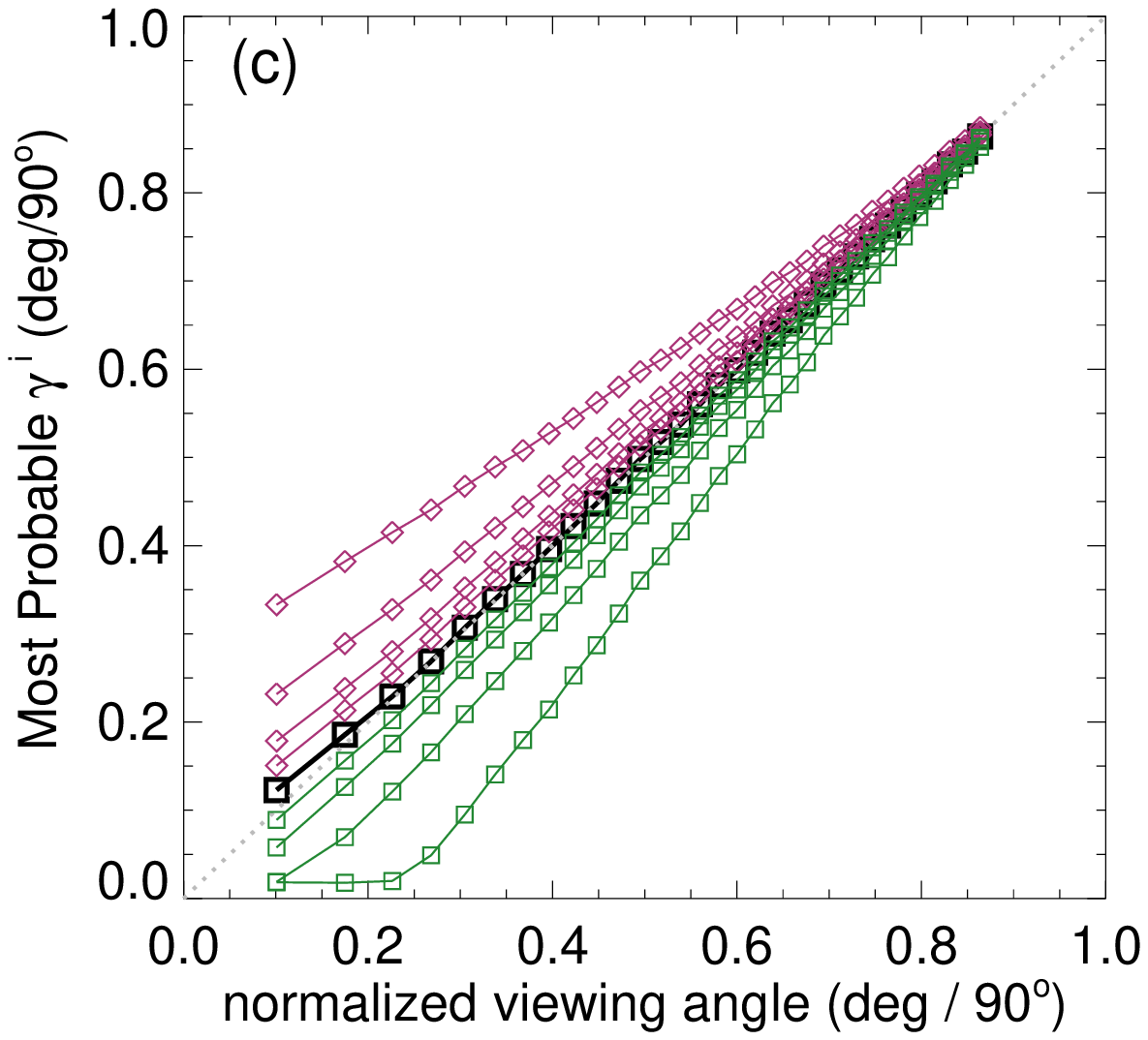}
\includegraphics[width=0.50\textwidth,clip, trim = 5mm 2mm 6mm 6mm, angle=0]{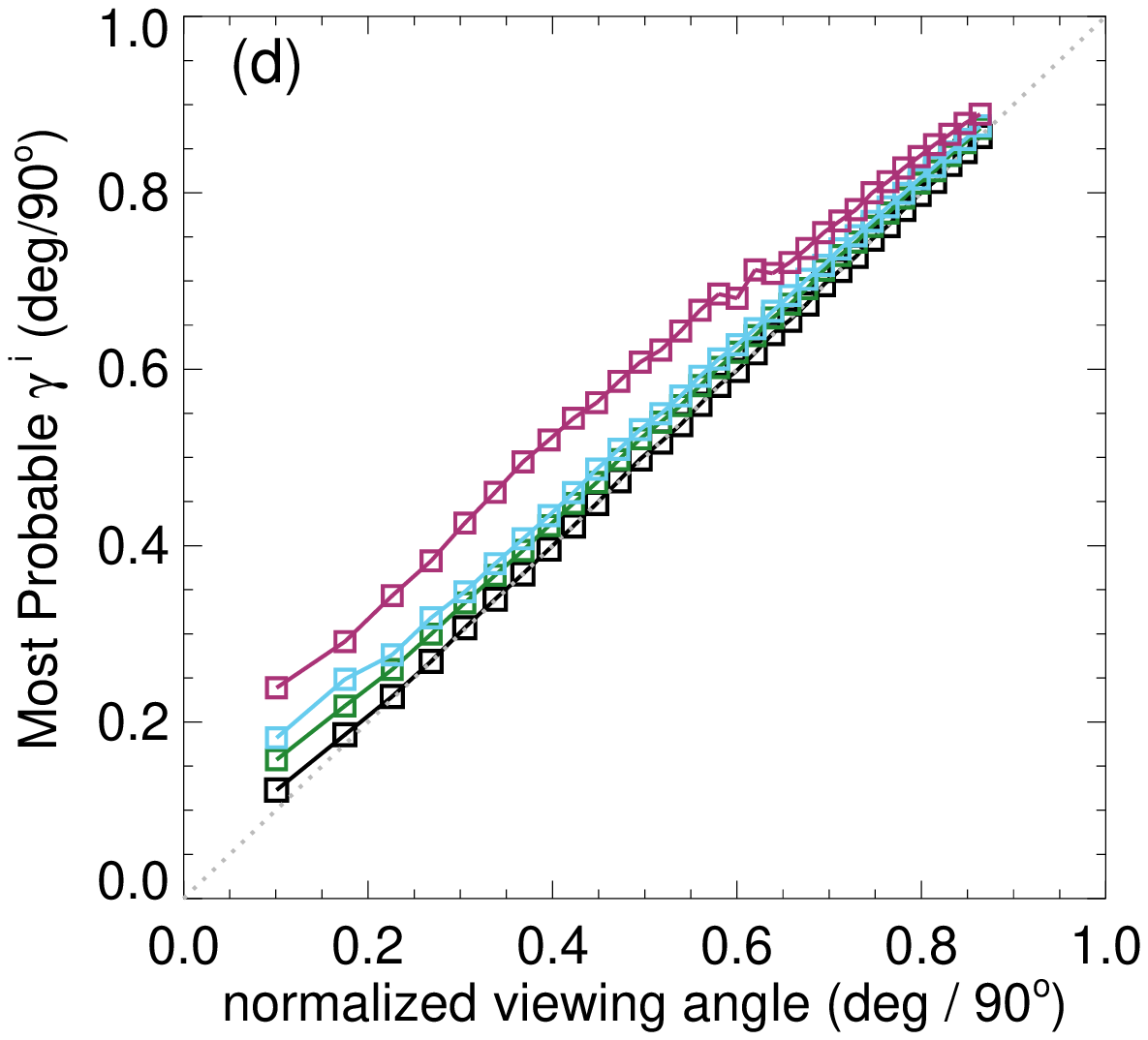}}
\caption{Results of model tests, demonstrating just one approach at a time and its behavior. 
For each, the black boxes are `no noise, no bias' results for reference.
(a) Experiment (1): no noise, constant $\Bperp$ bias at 
$\pm 50, 100, 200, 400 {\rm Mx\,cm}^{-2}$ (red diamonds/green squares); 
(b) Experiment (2): no noise, $\Bperp$ bias
at $\pm 10, 25, 50$\% (red diamonds/green squares) of total field; 
(c) Experiment (3): no noise, $\Bperp$ bias
is added at levels $\pm [50, 100, 200, 400]*\mu$ (red diamonds/green squares); 
(d) Magnetogram noise
only is added (no additional bias) for $\Bpar, \Bperp$ respectively at the $[0,0]$ ('no noise', black),
$[5, 100]$ ('low noise', green), $[10,200]$ ('medium noise', light blue) and $[50,300]$ ('high noise', red) levels (see text).  For all, the $x=y$ line is indicated.}
\label{fig:model_results}
\end{figure}

\begin{figure}
\centerline{
\includegraphics[width=0.33\textwidth,clip, trim = 6mm 2mm 7mm 5mm, angle=0]{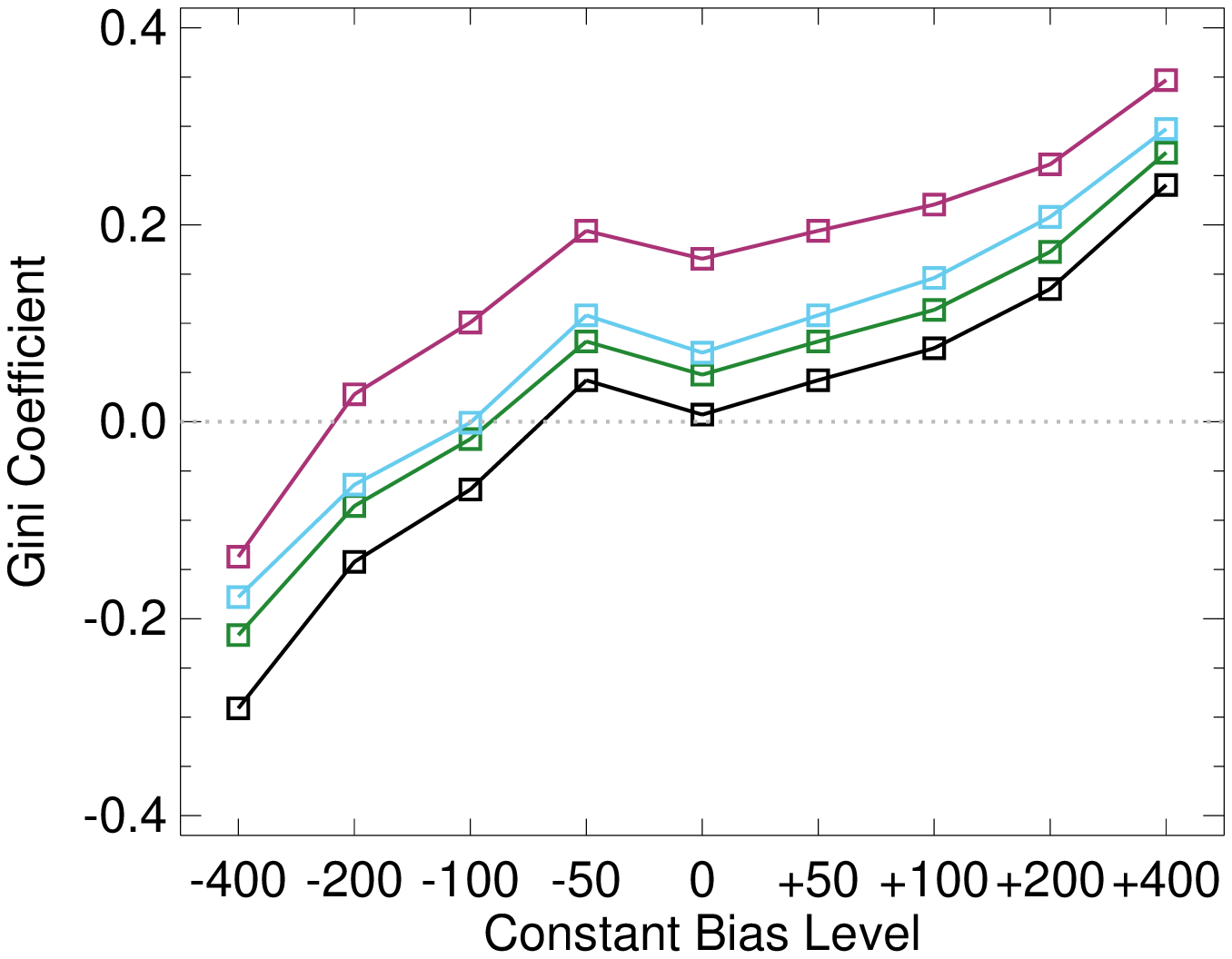}
\includegraphics[width=0.33\textwidth,clip, trim = 6mm 2mm 7mm 5mm, angle=0]{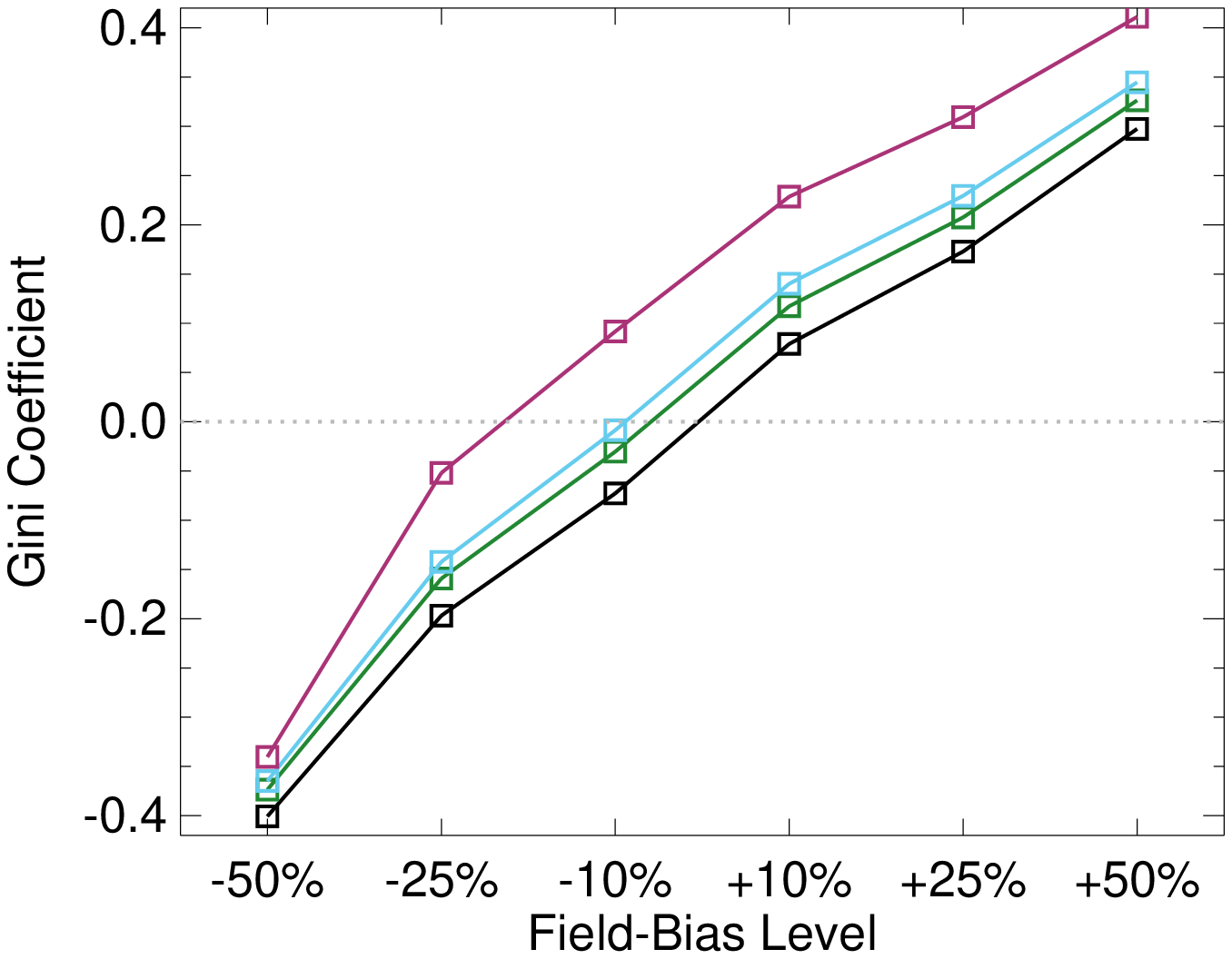}
\includegraphics[width=0.33\textwidth,clip, trim = 6mm 2mm 7mm 5mm, angle=0]{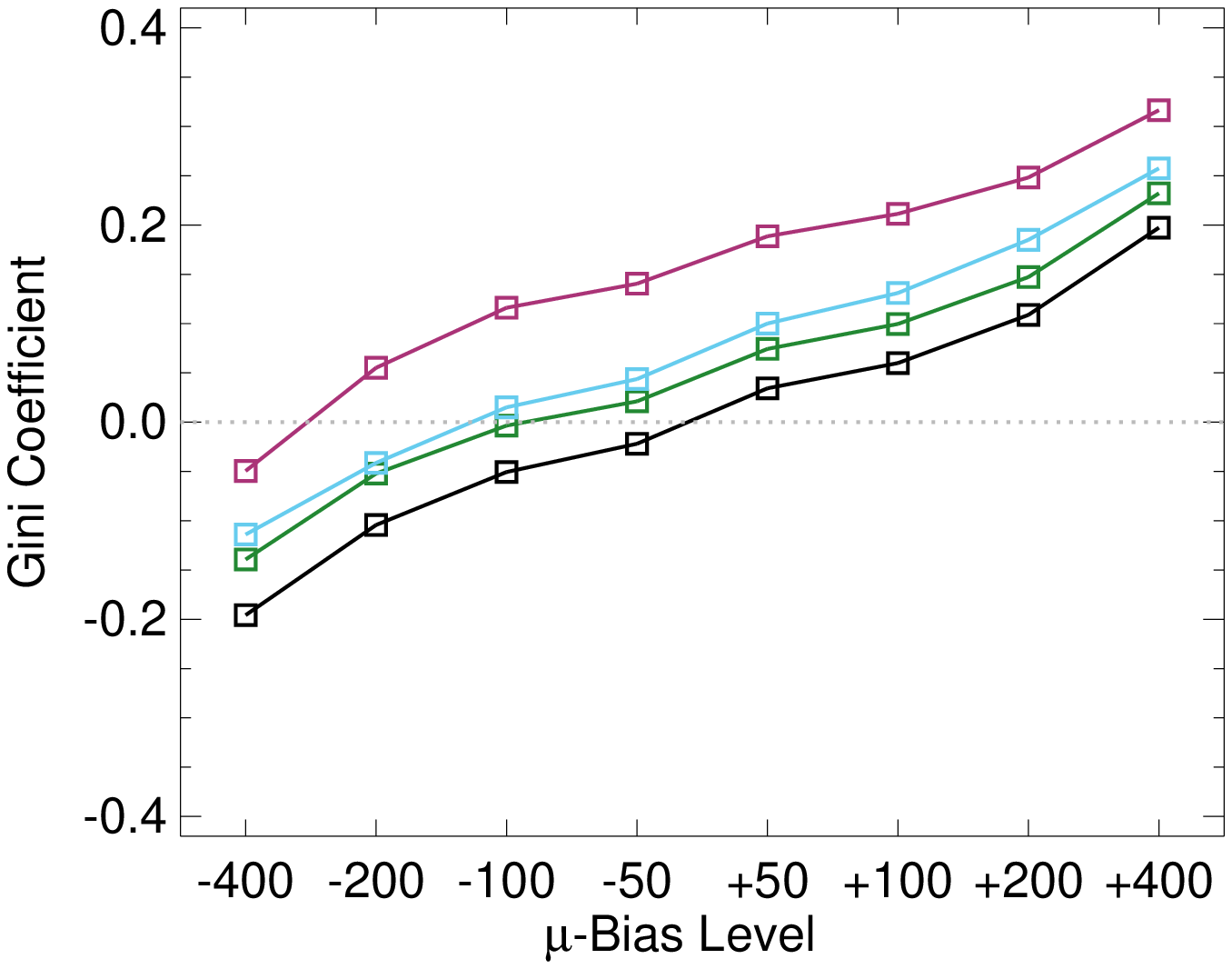}}

\caption{Summary statistics (Gini coefficients $\mathcal{G}$) for the
toy-model experiments, here including the behavior of the cross-purpose
noise and bias combinations ({\it e.g.} adding bias and photon noise,
both).  Left: Experiment (1): ``constant Bias Level'' applied, Middle:
Experiment (2): the bias as a fraction of the total field strength, Right:
Experiment (3): bias level that is then multiplied by $\mu=\cos(\theta)$
for a model of the effect of mis-representing $f\!f$.  Colors indicate
photon noise levels: Black: ``No Noise'', Green: ``Low Noise'', Light
Blue: ``Mid Noise'', Red: ``High Noise''.}

\label{fig:model_stats}
\end{figure}

The behaviors of the ``observed'' $E(|\gamma^i|)$ with viewing
angle as per experiments with bias are shown in
Figure~\ref{fig:model_results} (panels (a)-(c) and for photon 
noise only (panel (d)).  The `optimal' situation of no photon
noise and no bias is the $x=y$ line.  Similar to above, the addition
of bias produces deviations from the ideal case but for the bias cases
(panels (a)-(c)) the only unique signature is the spread at large viewing
angle, and a slight difference in shape. To summarize the performance
when both bias and noise are included, we present Figure~\ref{fig:model_stats} where
we focus on the Gini coefficient $\mathcal{G}$ as the bias levels are varied.  
The black ``no noise'' cases will reflect the curves in Figure~\ref{fig:model_results} (a)-(c),
the other curves are as labeled.  The impact of including both noise and bias is 
to increase the departure from $\mathcal{G}=0$ overall and bring
even negative bias toward $\mathcal{G}>0$.  The latter effect is caused by effectively a 
``canceling out'' effect between the bias and the noise.  Given the noise 
levels in {\it SDO}/HMI data are roughly between the ``Mid Noise'' and ``High Noise'' cases,
we can possibly use these plots to rule out some extreme cases, but the curves
and behaviors are similar enough to probably preclude 
determining a functional form of the bias.

We find that the impacts of over- or under-estimating $\Bperp$ do not always manifest
symmetrically about the $x=y$ line for the $E(|\gamma^i|)$ {\it vs.} $\mu$
relation.  And while the shapes of the curves deviate slightly from linear 
as both photon noise and bias are added, we are unable with this
toy model to replicate the curvature observed in the {\it SDO}/HMI pipeline output 
(Fig.~\ref{fig:gini_eg}), even when testing
a bias whose behavior is a function of 
observing angle, as expected by errors in fill fraction.  The $E(|\gamma^i|)$ {\it vs.} $\mu$
behavior of Spot \#2 is less curved, and closer to the curves produced by the toy model 
with moderate levels of positive bias.
The difference in the $E(|\gamma^i|)$ {\it vs.} $\mu$ behavior (curved {\it vs.} closer
to linear) between the Spot \#2 and the plage points in Figure~\ref{fig:gini_eg} is thus 
likely due to unresolved 
structures and hence non-unity $f\!f$ in the latter, and how that manifests with observing
angle.  Further model development to investigate the impact of unresolved structure
is beyond the scope of this paper \citep[although see the approach in][]{magres}.

\subsection{Results from Testing Mitigation Approaches}

\begin{figure}
\centerline{
\includegraphics[width=0.75\textwidth,clip, trim = 8mm 40mm 2mm 12mm, angle=0]{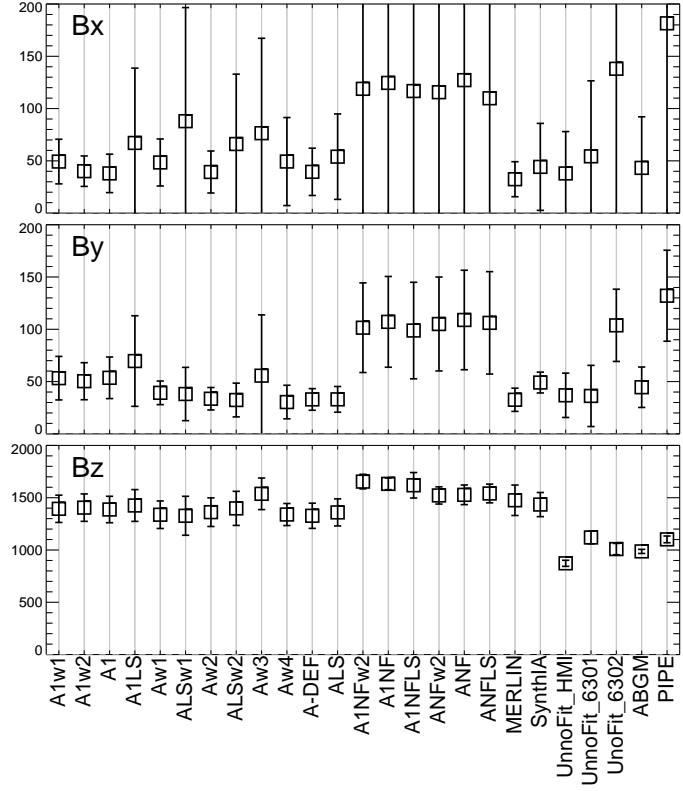}}
\caption{Boxes: the mean absolute separation of the most-probable values over the 6 days; error 
bars: the standard deviation of the {\it signed} separation, which will be larger
if it changes sign for example.  Results (top to bottom) are for $\Bxh, \Byh, \Bzh$ according
to the labels on the x-axis (introduced in Table\,\ref{tbl:Hinodegrid}).}
\label{fig:12457_mean_abs_sep}
\end{figure}

\begin{figure}
\centerline{
\includegraphics[width=0.5\textwidth,clip, trim = 8mm 40mm 2mm 12mm, angle=0]{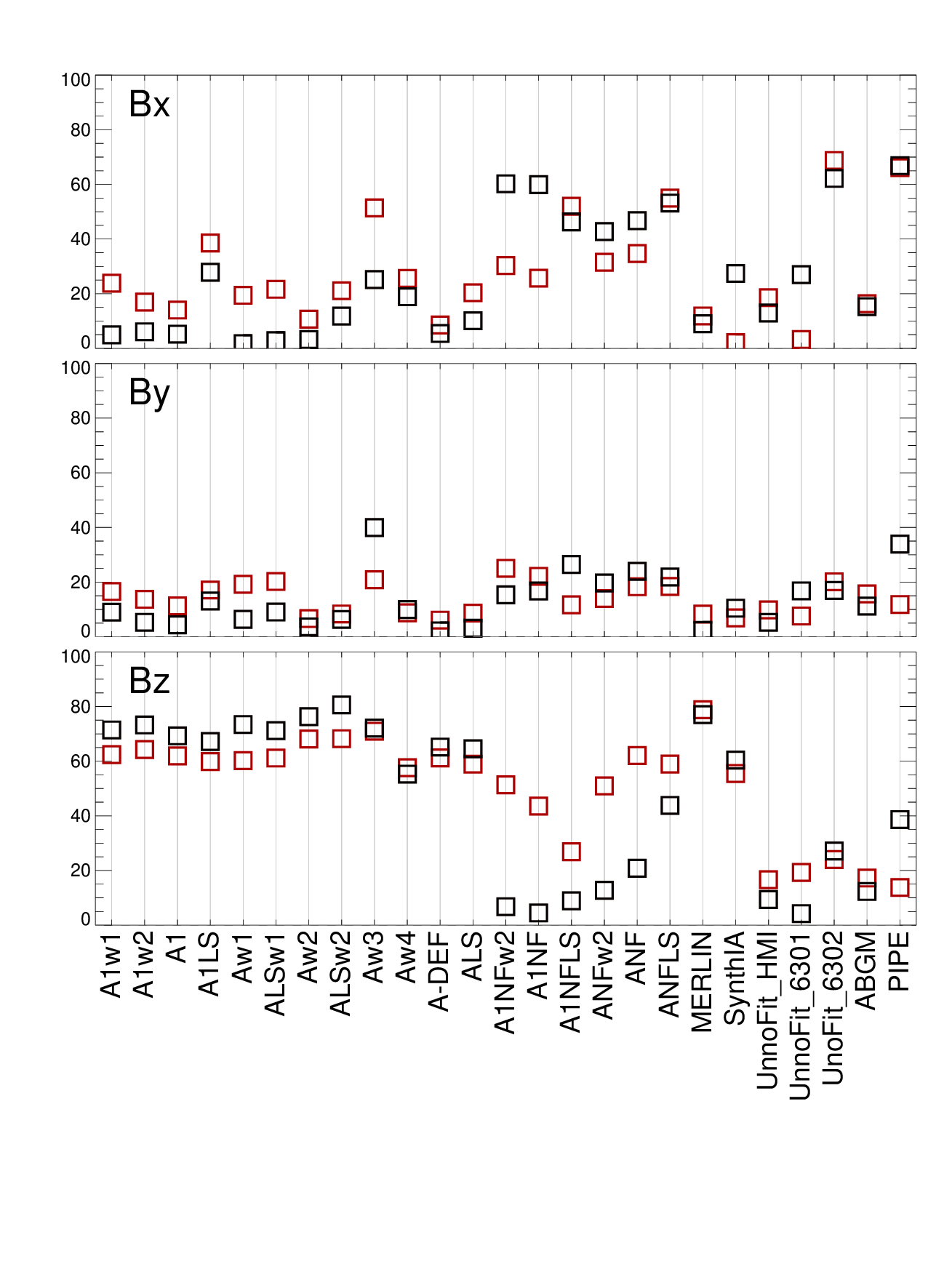}
\includegraphics[width=0.5\textwidth,clip, trim = 8mm 40mm 2mm 12mm, angle=0]{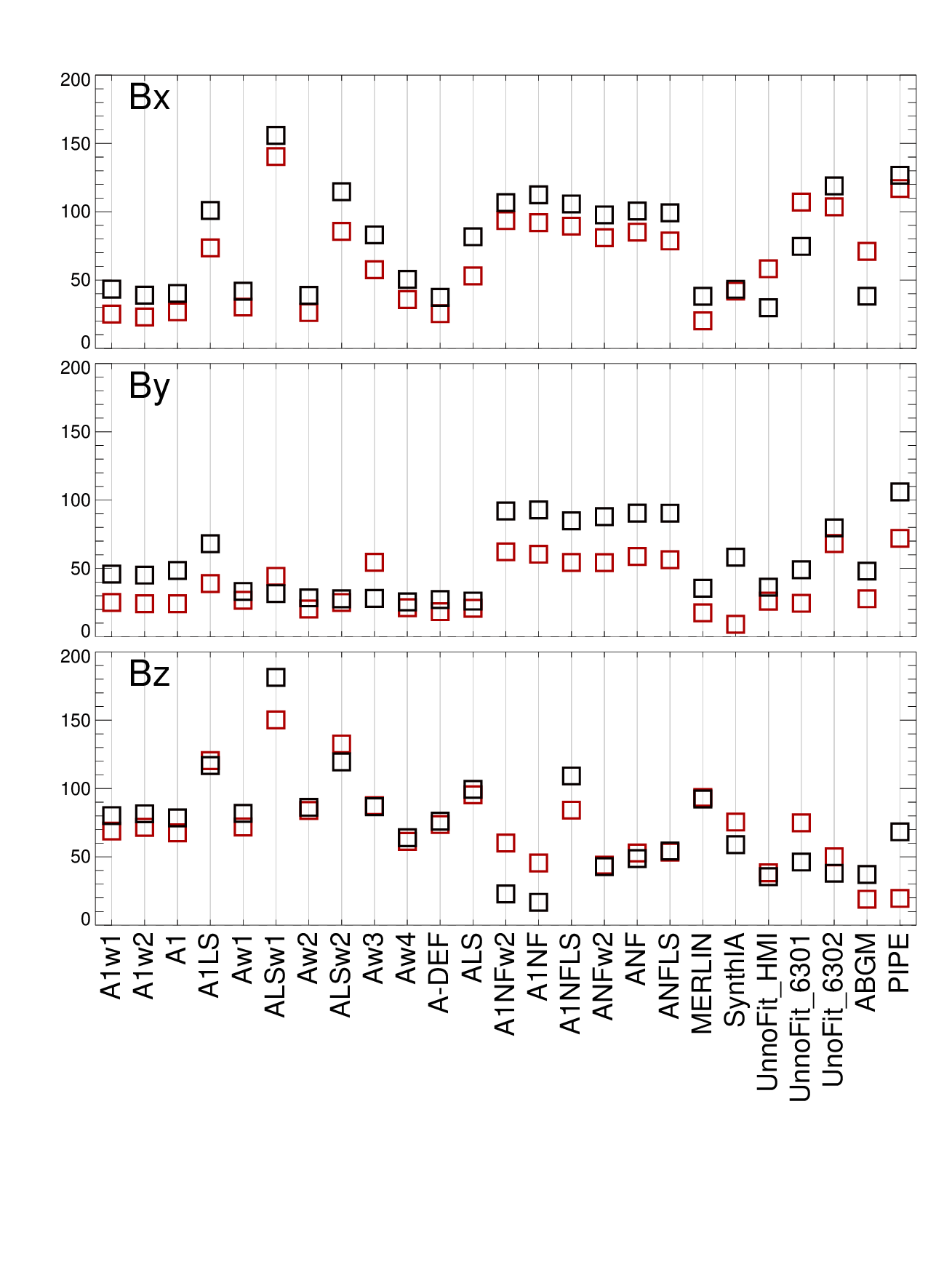}}
\caption{(Left) Median absolute deviation (MAD), and (right) maximum absolute deviation of the
most probable values over the 6 samples.  Red/Black points are for those with positive / negative
underlying $\Bzh$, respectively.}
\label{fig:12457_madmax}
\end{figure}

\begin{figure}
\centerline{
\includegraphics[width=0.75\textwidth,clip, trim = 8mm 176mm 2mm 12mm, angle=0]{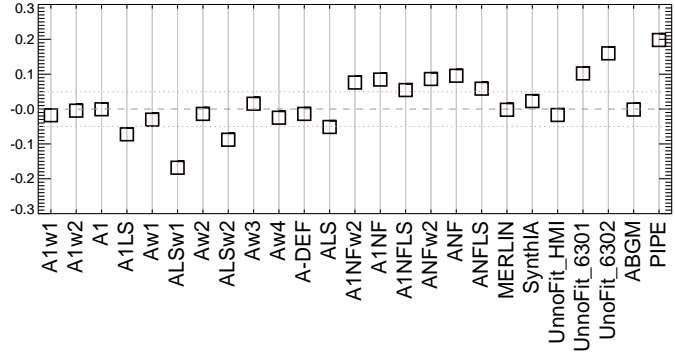}}
\caption{Gini coefficient of the distribution of $|\gamma^i|$ as $f(\mu)$, where the accumulated
6 maps were combined and then points binned according to $\mu$.  Gini coefficients within $\pm 0.05$
should show minimal bias.}
\label{fig:12457_gini}
\end{figure}

\begin{figure}
\centerline{
\includegraphics[width=0.75\textwidth,clip, trim = 8mm 26mm 3mm 5mm, angle=0]{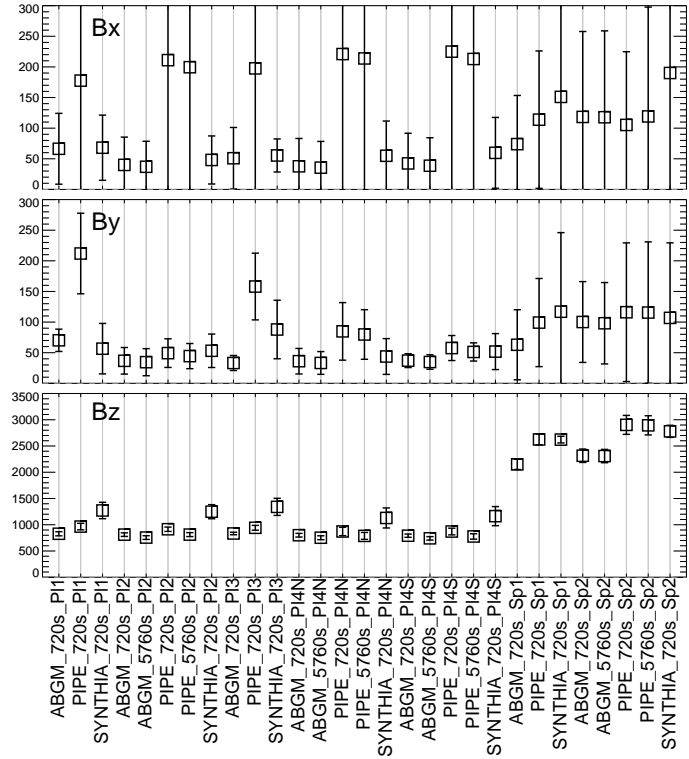}}
\caption{Similar format as Figure~\ref{fig:12457_mean_abs_sep} for the {\it SDO}/HMI target patches
and inversion options as indicated (labels are introduced in 
Tables\,\ref{tbl:Hinodegrid}, \ref{tbl:HMIgrid}, 
plus the {\tt hmi.S\_5760s}-series integration option), computed for
their disk transit.  Note that the sunspots are unipolar, which will influence the results.}
\label{fig:patches_mean_abs_sep}
\end{figure}

\begin{figure}
\centerline{
\includegraphics[width=0.5\textwidth,clip, trim = 10mm 26mm 3mm 6mm, angle=0]{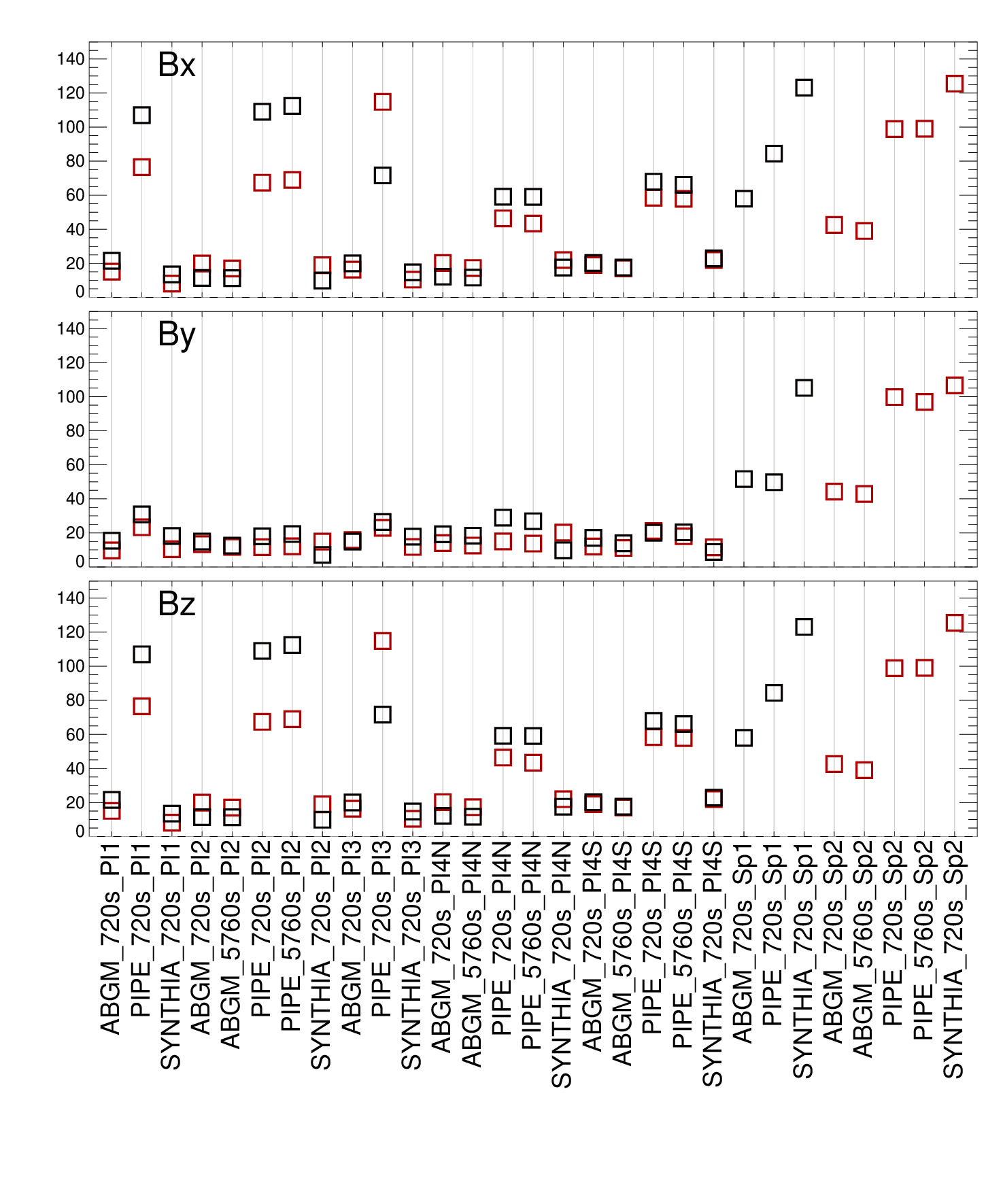}
\includegraphics[width=0.5\textwidth,clip, trim = 10mm 26mm 3mm 6mm, angle=0]{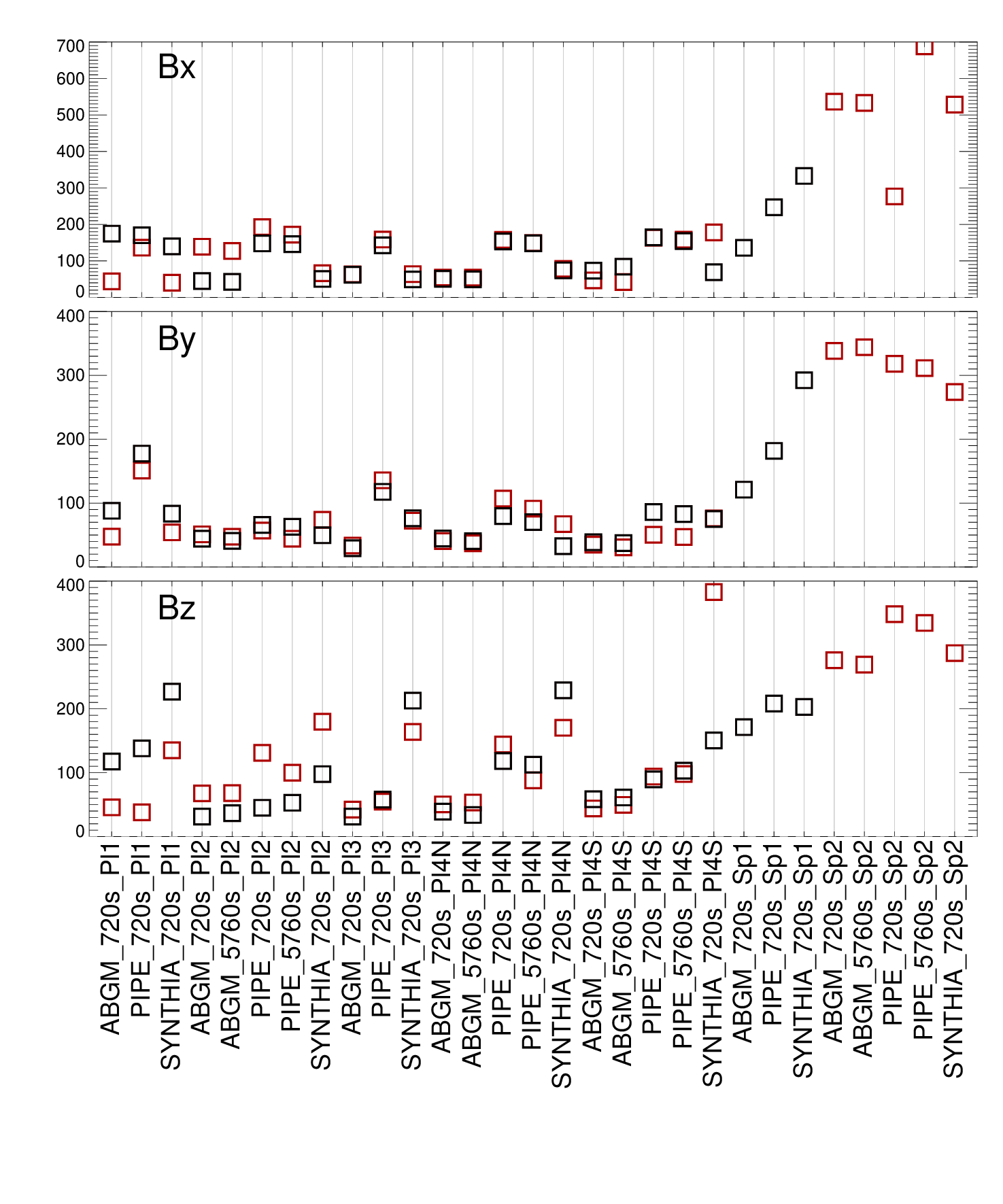}}
\caption{Following Figures~\ref{fig:12457_madmax} and \ref{fig:patches_mean_abs_sep},  
(Left) Median absolute deviation (MAD), and (right) maximum absolute deviation of the
most probable values over the $\approx 120$ samples as the regions transit the disk.  
Red/Black points are for those with positive / negative
underlying $\Bzh$, respectively; for the spots, there will be only one box for the dominant
sign.}
\label{fig:patches_madmax}
\end{figure}

As discussed in Section~\ref{sec:inversions}, there are other aspects of 
interpreting specropolarimetric signals that can lead to differences in 
inferred flux densities.  The results of our tests are summarized in 
Figures~\ref{fig:12457_mean_abs_sep},~\ref{fig:12457_madmax}, and \ref{fig:12457_gini}, 
and then in Figures~\ref{fig:patches_mean_abs_sep},~\ref{fig:patches_madmax}, 
and \ref{fig:patches_gini},
using the quantitative evaluation metrics on both image-plane $|\gamma^i|$ and 
heliographic $\Bxh, \Byh, \Bzh$ components (see 
Sections\,\ref{sec:helio_components_distrib},~\ref{sec:incl_distrib_metric}).  While the latter
indeed rely on the resolution of the $180^\circ$ ambiguity in the $\Bperp$
component, the same method was applied throughout, addressing one point
in \citet{SainzDalda2017}.  The $\mathcal{G}$ can
be positive or negative, all other metrics are positive; all 
metrics used here tend to zero for best performance (less bias).
For all metrics evaluated on time-series data, there may be
evolution or field-of-view departures that result in non-zero metrics, 
but all of the methods are evaluated on consistent data, so comparisons
between methods as grouped here are valid.

From the analysis of the AR\,12457 data we find that:
\begin{list}{$\circ$}{\topsep0cm\itemsep0cm}
\item Fitting multiple lines {\it vs.} only fitting one line in some experiments 
provided marginal improvement {\it c.f.} A1w1 {\it vs.} Aw1, A1w2 {\it vs.} Aw2
but which line was used and the data treatment also influence
the magnitude of the bias ({\it c.f.} the two \unno\ results).
\item \merlin\ and \asp\ default, or A-DEF, results are 
almost indistinguishable across the metrics, as expected given 
the heredity of the codes.
\item The bias is not just a matter of the quality of the input
spectropolarimetric data. Inversions of AR\,12457 using {\it SDO}/HMI input data 
that account for $f\!f$ (\synth, \vfisva, \unno)  
were by many metrics almost as good as the \merlin\ and \asp\ (default configuration
and others) results using {\it Hinode}/SP input spectral data. 
\item Optimization method can matter: generally, genetic-algorithm minimization 
performed slightly better than minimization by least-squares 
{\it c.f.} Aw1 {\it vs.} ALSw1, A-DEF {\it vs.} ALS, Aw1 {\it vs.} ALSw1, but
this is not a strong result in these tests.
\item Weighting differences can influence the results; equal weighting 
(Aw3) performed the worst ({\it c.f.} Aw3 {\it vs.} Aw1, Aw2, and Aw4).
\item Explicitly fitting for the fill fraction provides by far the most significant 
impact to reduce the bias, ({\it c.f.} ABGM {\it vs.} PIPE, A-DEF {\it vs.} ANF, 
ALS {\it vs.} ANFLS.
\end{list}

\noindent
From the analysis of the {\it SDO}/HMI time-series data (of both plage and spots),
we find that:

\begin{list}{$\circ$}{\topsep0cm\itemsep0cm}
\item The bias clearly manifests in all three local components of the field confirming
that bias in $\Bperp$ contaminates the determination of the true magnetic vector.  When
the impact of the $\Bperp$ bias is high ({\it e.g.} PIPE\_720s\_Pl1
{\it vs.} ABGM\_720s\_Pl1 results), it is generally high for all metrics across all three
$\Bxh, \Byh, \Bzh$ components.
\item The bias can manifest in strong / pixel-filling regions as well as 
unresolved features.  Focusing on *\_Sp1 and *\_Sp2, while we may expect 
some evolutionary changes with disk passage, the metrics improve
with inversions that fit for $f\!f$ ({\it c.f.} ABGM\_720S\_Sp1 {\it vs.} PIPE\_720S\_Sp1).
\item Reducing the photon noise (720s {\it vs.} 5760s) provides a small, but not 
significant mitigation.  Confirming the results of \citet{btrans_bias_1}, random photon 
noise in the data is not the primary source of $\Bperp$ bias.
\item Explicitly accounting for $f\!f \neq 1.0$ mitigates the bias, whether
through a traditional inversion approach ({\it c.f.} ABGM\_720S\_* {\it vs.} PIPE\_720S\_*) 
or by a Neural Net trained on inversion output that itself explicitly accounted 
for $f\!f \neq 1.0$ ({\it c.f.} SYNTHIA\_720S\_* {\it vs.} PIPE\_720S\_*).
\end{list}

\begin{figure}
\centerline{
\includegraphics[width=0.75\textwidth,clip, trim = 11mm 158mm 3mm 6mm, angle=0]{PAPER_PATCHES_GINI.eps}}
\caption{Following Figures~\ref{fig:12457_gini} and \ref{fig:patches_mean_abs_sep}, the Gini coefficient 
of the distribution of $|\gamma^i|$ as $f(\mu)$, as the patches sampled across
the disk.  Results within $\pm 0.05$ should show minimal bias.}
\label{fig:patches_gini}
\end{figure}

\section{Post-Facto Quick Fix}
\label{sec:qnd}

Using the assumption that plage is {\it statistically}
dominated by radially-directed field and the fill factor $f\!f$ is not 
explicitly set or is otherwise already multiplied-through in the equations below, then 
we use the most probable image-plane inclination
$\mathcal{P}(|\gamma^i|) = \theta$, $\theta$ being the observing angle
\citep{Svalgaard_etal_1978,WangSheeley1992,bbpot}, as the basis for evaluation
and now, for correction.  This ``quick-fix'' approach is straightforward
\citep[see also][]{RudenkoDmitrienko2018},
but should be implemented when concerned with the large-scale
averages or contributions across well-measured (sufficient polarization)
pixels.  The resulting values are not claimed to be ``correct'' any more
than any inversion results are, just less influenced by the bias.

The approach is as follows for under-resolved plage areas:

\begin{enumerate}
\item Identify plage structures sampled across a range of observing angles $\mu=\cos(\theta)$
(with full-disk data or with a temporal sequence of {\it statistically} consistent
structures.
\item Examine the distributions of $|\gamma^i|$ as $f(\mu=\cos(\theta))$ and
determine a simple functional form for the systematic angle difference 
$\Delta|\gamma^i| = \mathcal{P}(|\gamma^i|)(\mu) -\mu$ in appropriate units.
\item Assuming that $\Bpar$ should not change, and working initially with the absolute
inclination $|\gamma^i|$ and assigning a new inclination $|\gamma^i|^\prime = |\gamma^i| - \Delta|\gamma^i|$,
we find
\begin{eqnarray}
\Bpar  & = & |\BB| \cos(|\gamma^i|)  =  |\BB|^\prime \cos(|\gamma^i|^\prime) \\
|\BB|^\prime &=& |\BB| \cos(|\gamma^i|^\prime) /  \cos(|\gamma^i|) \\
 &=& \frac{|\BB|}{\cos(\Delta|\gamma^i|) + \sin(|\gamma^i|)\sin(\Delta|\gamma^i|)/\cos(\Delta|\gamma^i|)}
\end{eqnarray}
where the $\Delta\gamma^i$ is the functional form of the deviation of
the most probable image-frame inclination from expected as a function of
observing angle, and each pixel is thus ``corrected'' according to 
field strength, inclination angle, this function, and the observing
location.  This formulation can of course be presented in a number
of ways, but the important aspects are that (1) if $\Delta|\gamma^i| \rightarrow 0$
that $|\BB|$ is recovered, and (2) the results do not become infinite
for, {\it e.g.}, $|\gamma^i|\approx0$ (if it performs badly near $|\gamma^i|\approx \pi/2.$ that
will statistically be less of an issue).

Once the total field strength $|\BB|^\prime$ is adjusted in this way, we work backwards
to find the new inclination angle and the new components:
\begin{eqnarray}
\gamma^{i\prime} &=& \cos^{-1}(\Bpar / |\BB|^\prime) \\
\Bperp^\prime &=& |\BB|^\prime \sin(\gamma^{i\prime})
\end{eqnarray}
where note that in the last two steps, we now use and recover the full $[0,\pi]$ range
of $\gamma^i$.  Again, $\Bpar$ remains the same.
\end{enumerate}

\begin{figure}
\centerline{
\includegraphics[width=0.65\textwidth,clip, trim = 0mm 0mm 13mm 0mm, angle=0]{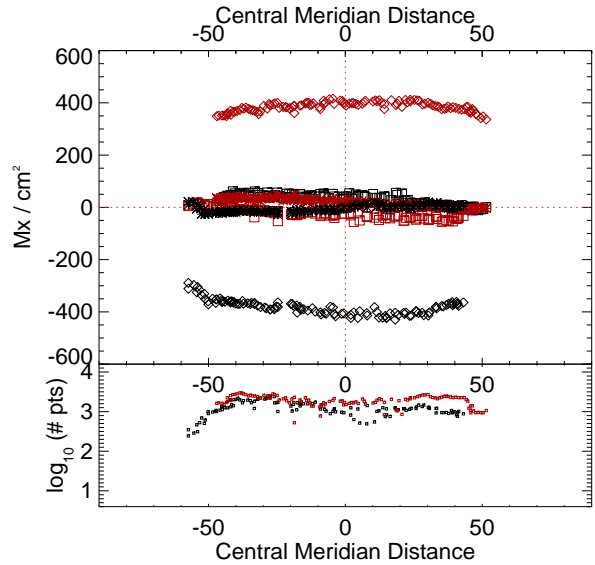}}
\caption{Similar to Figure~\ref{fig:vpipe_ts_plage3} but after the corrections
described in the text.   The metrics, for $\Bxh,\ \Byh\ {\rm and}\ \Bzh$
respectively are: Mean Absolute Separation: $32.0\pm25.2,\ 57.3\pm 16.6,\ 765.6\pm 42.5$;
MAD (for $\Bzh>0\,/\, \Bzh<0$): $13.5/9.4, 14.7/12.1, 12.7/19.0$; 
MaxAD: $42.4/26.0, 53.2/57.9, 42.5/80.4$; Gini: $-0.02, -0.02$.
}
\label{fig:post_facto}
\end{figure}

As a demonstration, we find the coefficients of a $2^{\rm nd}$-order
polynomial fit for $\Delta|\gamma^i|$ to $\mu$ from full-disk {\it
SDO}/HMI data for 2010.04, 2010.06 and 2010.08.  A $3^{\rm rd}$-order
fit the sampled regions well, but extrapolated poorly for disk-center
corrections.  We then apply the above to the Plage 3 time-series
extraction (Figures\,\ref{fig:hmi_plage3},\,\ref{fig:vpipe_ts_plage3}).
A first try resulted in an over-correction, diagnosed
using a most-probable $|\gamma^i|(\mu)$ plot similar to
Figure~\ref{fig:gini_eg}, but a simple 25\% reduction in the
coefficient magnitudes produced a narrow distribution centered well
on the $x=y$ line.  The disambiguation was then performed and the
same diagnostics are used to evaluate this {\it post-facto} approach.
The relevant metrics {\it c.f.} the results for ``PIPE\_720s\_Pl3''
in Figures\,\ref{fig:patches_mean_abs_sep}-\ref{fig:patches_gini}, are
listed in the caption for Figure\,\ref{fig:post_facto} where we show
the new time-series plots to compare to Figure\,\ref{fig:vpipe_ts_plage3}.

\section{Concluding Remarks}

The interpretation of spectropolarimetric signals in terms of the 
magnitude, orientation, behavior, and location of magnetic fields
in the solar atmosphere is simply not straightforward.  How to 
approach the problem depends severely on the scientific question
being posed.  With regards to interpreting the large-scale vector field 
of under-resolved features, it is now well-established that 
how the spectropolarimetric signals are handled can contribute
to a systematic bias in the component perpendicular to the line-of-sight, 
$\Bperp$.  

Previous works on this topic have illustrated that the problem exists,
and touched on some compute-intensive approaches to mitigation.  In 
this work we perform four novel investigations and point to 
potentially less-arduous mitigation approaches.

First, we develop quantitative metrics to measure the $\Bperp$ bias.
At least one, the Gini-coefficient of the departure from expected
image-plane inclination angles in plage features, is applicable without
performing the $180^\circ$ disambiguation, given sufficient sampling
of appropriate structures across observing angle.  These quantitative
metrics are used to point out that the bias in {\it SDO}/HMI data is
not limited to the unresolved plage features; some bias is seen in the
strong-field field-filling pixels of sunspots.  The quantitative metrics
are then available to evaluate and compare data.

Second, we systematically compared the results of different inversion
implementation options and targets (different spectral lines used),
evaluating the results using common observational targets and the
quantitative metrics developed above.  We find that the most impactful
implementation choice is to include $f\!f$ as an independent parameter
in the optimization, or having trained a neural net on such data.
Applying inversions that explicitly fit $f\!f$ mitigates bias in plage
{\it and in strong-field pixel-filled sunspot centers}.  That being
said, which optimization scheme is used, the weighting used to calculate
$\chi^2$, and the use of multiple spectral lines also impact the outcomes.
We also tested a direct comparison of two very different data sources
({\it Hinode}/SP {\it vs.} {\it SDO}/HMI for NOAA\,AR\,12457) using
different inversions both, and found that data source {\it per se}
did not account for the bias.

Third, we construct a simple ``toy'' model that is appropriate to test
certain observed features of the $\Bperp$ bias.  
Experiments were performed which added bias with different functional forms
as well as varying the level of photon noise in the data.
We show that noise is not the primary contributor
to the of the $\Bperp$ bias. We could not, however, reconstruct a distinct non-linear
aspect visible in the {\it SDO}/HMI $|\gamma^i|(\mu)$ behavior. 
The source and contribution function to the bias is more complex than 
our experiments.  Further refinement to the models to determine
the source of this non-linear shape is beyond the scope of this paper.

Fourth and finally, we demonstrate a straightforward ``quick fix'' that
can be applied for analysis of global plage structures on a statistical
basis, and can be applied prior to performing the disambiguation step.
This {\it post-facto} algorithm 
\citep[see also][]{RudenkoDmitrienko2018},
should not take the place of a more
robust inversion, and should probably not be used to interpret any individual
pixel's results quantitatively.  However, it can be used to produce
vector field data for which the bias is adequately removed so as to 
produce more reasonable global $\Bph,\ \Bth,\ \Br$ maps for example.

We show that there are in fact viable options for more robust full-disk 
inversions of {\it SDO}/HMI data:  \unno\ \citep{unnofit}, the new \vfisva\ \citep{abgm_etal_2021}
and \synth\ \citep{synthia}.  The first two are more traditional implementations
of Milne-Eddington codes that explicitly include fill fraction in the
optimization, the latter is a neural-net trained on {\it Hinode}/SP
Level-2.0 output and {\it SDO}/HMI {\tt hmi.s\_720s} $[I,~Q,~U,~V]$ 
input.  Given the vast differential in computing resources required,
the latter may be a more readily-available solution, especially for large datasets.

\section*{Data Availability Statement}
The majority of the data used in this research is sourced from public
pipeline-produced products and codes, as cited and referenced; specific
time-series cubes, for example, can be made available upon request to the corresponding
author.  Data from 
\vfisva\ are in series {\tt su\_abgm.ME\_720s\_fd10\_forKD2} at {\url{http://jsoc2.stanford.edu/ajax/lookdata.html}} which requires interested parties to contact the {\it SDO}/HMI team for access.
Data from \synth\ can be made available upon reasonable request, see also
\citet{synthia} and sources therein; KDL, ELW, and RH 
are actively involved in a project to establish the \synth\ 
as a publicly-accessible series through {\tt JSOC}.  The \unno\ 
inversion code is available at {\url{http://lesia.obspm.fr/perso/veronique-bommier}}.

\begin{acks}
We thank the anonymous referee for very helpful suggestions.
KDL and ELW gratefully acknowledge support from NASA GSFC 80NSSC19K0317,  
NASA/LWS 80NSSC18K0180, 
from Lockheed-Martin Space Systems contract \#4103056734 for Solar-B FPP Phase E, 
and NASA/HSR 80NSSC18K0065 for developing the custom-extraction 
codes. 
We thank Dr. Yang Liu and Dr. Keiji Hayashi for helping identify suitable
target regions, Dr. Xudong Sun for running 5760\,s {\it SDO}/HMI pipeline 
data, Drs. Ale Pacini and Graham Barnes for helpful discussions, and Dr. David Fouhey for 
nudging RH, who (with KDL) acknowledges support from the NASA Heliophysics Phase-I DRIVE 
Science Center (SOLSTICE) at the University of Michigan under grant
NASA 80NSSC20K0600; RH's work was also supported by the Michigan Institute for Data Science via 
a Propelling Original Data Science grant.
ABGM acknowledges support by NASA contract NAS5-02139 (HMI) to Stanford University, and 
additional support by the Research Council of Norway through its Centres of Excellence 
Scheme, project No.  262622. 
VB acknowledges the access granted to 
the HPC resources of MesoPSL financed
by the Region Ile de France and the project Equip@Meso (reference
ANR-10-EQPX-29-01) of the programme Investissements d’Avenir supervised
by the Agence Nationale pour la Recherche.
Some figures in this manuscript were produced using IDL colour-blind-friendly colour 
tables \citep[see][]{pjwright}.
\end{acks}

\bibliographystyle{spr-mp-sola}
\bibliography{ms_arxiv.bib}  

\end{article} 
\end{document}